\documentclass[a4paper,11pt]{article}
\usepackage{jheppub} 

\usepackage[T1]{fontenc} 
 \usepackage{bm}
 \usepackage{ulem}
 \usepackage{amssymb,amsmath,epsfig,bm,pifont,amsfonts}
\usepackage[utf8]{inputenc}
\newcommand{\Ds}{\displaystyle}                           
\newcommand{\be}{\begin{equation}}\newcommand{\ee}{\end{equation}}%
\newcommand{\ba}{\begin{eqnarray}}                        
 \newcommand{\ea}{\end{eqnarray}}                         
\def\MSbar{\relax\ifmmode\overline                        
           {\rm MS}\else{$\overline{\rm MS}${ }}\fi}     
  \def\ie{\hbox{\it i.e.}{ }} \def\etc{\hbox{\it etc.}{ }}
   \def\eg{\hbox{\it e.g.}{ }} 

\newcommand{\dAA}{d_A^{abcd}d_A^{abcd}}
\newcommand{\dRR}{d_F^{abcd}d_F^{abcd}}
\newcommand{\dRA}{d_F^{abcd}d_A^{abcd}}                                           
\newcommand{\dRRNA}{\frac{d_F^{abcd}d_F^{abcd}}{N_A}}
\newcommand{\dRANA}{\frac{d_F^{abcd}d_A^{abcd}}{N_A}}
\newcommand{\dAANA}{\frac{d_A^{abcd}d_A^{abcd}}{N_A}}

\usepackage{color}
\definecolor{mBlue}{rgb}{0,0,1}
 
\definecolor{mRed}{rgb}{1,0,0}
 
\definecolor{mGreen}{rgb}{0.0,1.0,1.0}
 
    \definecolor{DarkGreen}{rgb}{0.04,0.5,0.1}

\newcommand{\black}[1]{{\color{black} #1}}                          
\newcommand{\red}[1]{{\color{red} #1}}
\newcommand{\blue}[1]{{\color{blue} #1}}
\definecolor{green}{rgb}{0.133,0.56,0}

\definecolor{DarkGreen}{rgb}{0.04,0.5,0.1}
\definecolor{GrayW}{rgb}{0.50196,0.50196,0.50196}
 
 \definecolor{amethyst}{rgb}{0.57, 0.36, 0.51}
  \newcommand{\amethyst}[1]{{\color{amethyst} #1}}

  \newcommand{\BluTn}[1]{\textcolor{blue}{#1}}
   
\def\1{\hbox{{1}\kern-.25em\hbox{l}}}

 \date{\today}

\title{
Optimization of perturbation series in QCD for physical quantities using
the renormalization group: necessary conditions and partial results
}

\author[a]{D. Kotlorz,}
\author[a]{S. V. Mikhailov}
 \affiliation[a]{Bogoliubov Laboratory of Theoretical Physics, JINR,
                141980 Dubna, Russia}
\emailAdd{dorota@theor.jinr.ru}
\emailAdd{mikhs@theor.jinr.ru}
\abstract{
We explore approaches to numerically optimize a segment of the
perturbative series for physical quantities using the QCD
renormalization group. We apply these methods to the
perturbative series for the coefficient function $C_{Bjps}$ of
the Bjorken polarized sum rule and the Adler function
$D_A$. 
Using various techniques proposed in the literature, we discuss
the consequences of ``optimization.''}

\begin{document}
\maketitle

\section{Introduction}
\label{sec:intro}
High-order calculations in quantum chromodynamics (QCD) are
extremely complex and require significant effort, so it
is crucial to use their results as efficiently as possible
through optimization. Here we discuss an approach to optimizing
truncated perturbative QCD (pQCD) series based on the
renormalization group (RG). The RG remains one of the most
pivotal tools for analyzing the results of QCD.
The RG equations define the running of the QCD coupling constant $\alpha_s$
and the the multiplicative transformation of matrix elements
with a change in the renormalization scale $\mu^2$.
We consider the case of renormalization group invariant (RGI) and
single-scale quantities -- depending on $Q^2$ or $s$ -- which have no
anomalous dimensions.
Therefore, the RG running reduces only to the change of QCD coupling
$\alpha_s \to \alpha'_s$ and the corresponding changes of the expansion
coefficients $c_n \to c'_n$.
The general scheme for the parametrization of the perturbation series
for RGI-quantities is given in the next section.
The crucial question is how to manage with the perturbation series
for the RGI quantity by varying the renormalization scale $\mu^2 \to \mu'^2$.
Our further investigations are more methodical in nature
rather than focused on optimizing perturbation series for specific RGI.
Their subjects are outlined in Items (1)-(3) below.

(1) First, we find that the variation of the scale $\mu^2$ should be
constrained in some way to preserve the validity of the perturbation
theory (PT) we use.
This was considered for the example of the coefficient function of the Bjorken
polarized sum rule ($C_\text{Bjp}$) in \cite{Kotlorz:2018bxp}.
In Section~\ref{sec:sec3}, we develop this approach for the admissible
values of $\mu^2$ for the RGI quantities $C_\text{Bjp}(Q^2)$ in the
leading twist and the nonsinglet Adler $D_\text{NS}(Q^2)$ function
\cite{Baikov:2008jh,Baikov:2010je,Baikov:2012zm}.
In this way, we explicitly define constraints on ``the playing field''
for analyzing  different transformations
and optimizations.

(2) Based on the above constraints, we consider a type of optimization
that takes the perturbation coefficients $c_n$ as a whole, without
initially dividing them into meaningful subparts.
This approach is discussed in Section~\ref{sec:sec4},
 where we apply a trick to eliminate some expansion coefficients of the series and improve convergence.
 Additionally, we investigate the extrema of the sum of radiative corrections,
$\sum_n a_s'^n c_n'$, where $a_s = \alpha_s / 4\pi$,
at various conditions.

(3) An alternative consideration suggests dividing the PT expansion
coefficients of RGI $c_n$ into two parts at the default value of the
scale $\mu^2 = Q^2$: $c_n=c^\text{rest}_n +\bar{c}_n$.
The first part $c^\text{rest}_n$ is fixed following outer requirements
of optimization.
Each of the second terms $\bar{c}_n$ is transformed to the new scale
$\mu'$, which defines a new coupling constant
$\alpha'_s = \bar{\alpha}(\mu')$.
The PT coefficients then transform as
$c_n \rightarrow c'_n=c^\text{rest}_n$.
It is expected that the new series with the terms
$(a_s')^n c^\text{rest}_n$ becomes in a sense more appropriate
than the initial one at $\mu^2=Q^2$.
This approach is traced from the well-known BLM method for the NLO case,
as described in \cite{Brodsky:1982gc}. In this method, the coefficient
$c_2$ is unambiguously decomposed into two parts:
$c_2 =c^\text{rest}_2+\beta_0\,c_2[1]$.
The element $c_2[1]$ is then transferred to a new scale
$\mu'=\mu \exp(-c_2[1]/2)$.
This value, $\mu'^2$, can be interpreted as a characteristic flow of
the pulse and, therefore, fixes the argument of $\alpha_s(\mu'^2)$ at
the ``true scale $\mu'^2$'', which is now different from the external scale
$\mu^2$ taken previously as $Q^2$ by default.

It is worth discussing the generalization of Item (3) in light of Item (1),
which is the subject of  Section\,\ref{sec:sec5}.
To generalize the BLM approach to higher orders $n>2$ of perturbation
theory, one needs to obtain explicit manifestations of the running of
$a_s$ in each coefficient $c_n$. For this classification,
the $\{\beta \}$-expansion was suggested, which allows us to consider
the effect of running of the intrinsic charge in full detail
\cite{Mikhailov:2004iq,Kataev:2014jba,Baikov:2022zvq,Mikhailov:2024mrs}.
In short, one divides the coefficients $c_n$ at each order $n$ in pQCD
into parts responsible for specific kinds of renormalization of the
intrinsic $\alpha_s(\mu)$ inherent in this RGI \cite{Mikhailov:2004iq}.
Those parts are proportional to the products
of the $\beta$-function coefficients $\beta_j$ that are applicable for this perturbation order
$n$, as discussed in Appendix \ref{App:A1}.
At any order $n$, there is one element $c_n[0]$ in the decomposition that
is not related to the charge renormalization (the so-called ``conformal''
element \cite{Mojaza:2012mf}), and it is not connected to any of the $\beta_j$.
This $c_n[0]$ and its interpretation play a key role in the ``Principle of
Maximum conformality'' (PMC) approach, which uses the condition
$c^\text{rest}_n = c_n[0]$.
The approaches that elaborate different types of $\{\beta \}$-expansion
are presented in the review \cite{DiGiustino:2023jiq} and
references therein.
Other types of such decompositions are discussed in \cite{Cvetic:2016rot}
and \cite{Kataev:2022iqf}, which are based on the Crewther-Broadhurst-Kataev
(CBK) relation \cite{Broadhurst:1993ru}.
Our findings are summarized in the conclusions section.

\section{Parametrization  of perturbation series for RG-invariant quantities} 
\label{sec:sec2}
We consider the transformation of the PT coefficients $c_i$ of an RGI observable $C$, which can be the coefficient function $C_\text{Bjp}(a_s)$ or the Adler function $D_A(a_s)$ ($R_{(e^+e^- \to h)}$) under the change in the normalization scale $\mu\to\mu'$.
  The perturbation expansion for the one-scale $C(Q^2, a_s(\mu^2))$ reads:
  \ba
\!\!\!\!\! C\left(\frac{Q^2}{\mu^2},a_s(\mu^2)\right)&=&
1+c_1^\text{ns}\Big(1\,a_s(\mu^2)+c_2~a_s^2(\mu^2)+ c_3~a_s^3(\mu^2)+c_4~a_s^4(\mu^2)+\ldots \Big)\label{eq:4.1b}\,,
\ea
where the coefficients $c_i=c_i\left(\ln(Q^2/\mu^2)\right)$ are calculated in the \MSbar scheme and
 are normalized by the first coefficient $c_1^\text{ns}$.
 Applying the $\ln(Q^2/\mu^2)$ as the argument of $c_i$, we emphasize that only logarithmic corrections can appear in the pQCD used.
 For $C_\text{Bjp}$ the $c_1^\text{ns}=-3C_\text{F}=-4$, while for $D_A$, or $R_{(e^+e^- \to h)}$ the $c_1=+3C_\text{F}=+4$; the running QCD coupling
$a_s$ is $ a_s(\mu^2) =\alpha_s(\mu^2)/(4\pi)$.
For the default  condition  $\mu^2=Q^2$ the coefficients $c_i\equiv c_i(0)$ are the numbers
presented in Appendix~\ref{App:A} in different forms.
Below we develop the approach suggested in \cite{Mikhailov:2004iq,Kataev:2014jba} and \cite{Kotlorz:2018bxp}, therefore, we are doomed here to some repetitions from ref. \cite{Kotlorz:2018bxp}.

\subsection{Parametrization of the RG transformation}
\label{sec:sec4a}

Let $a_s=\bar{a}_s(t)$ and $a'_s=\bar{a}_s(t')$ be the solutions of the RG equation for the QCD charge with
logarithmic argument $t=\ln(\mu^2/\Lambda_{qcd}^2)$ at the same integration constant $\Lambda_{qcd}^2$.
Re-expanding the running coupling $\bar{a}_s(t)= a_s(\Delta, a_s')$
in terms of  $\Delta=t-t'=\ln\left(\mu^2/\mu'^2\right)$ and the coupling $a_s'$, one obtains
\ba
\label{eq:RGrearrange}
a_s = a_s(\Delta, a_s')&\!=\!&\!
\exp\left[-\Delta \beta(\bar{a}_s)\partial_{\bar{a}_s}\right]\bar{a}_s\Bigr|_{\bar{a}_s=a_s'} \nonumber \\
&\!=\!& a'_s - \frac{\Delta}{1!} \beta(a'_s) +
\frac{\Delta}{2!}\beta(a'_s)\partial_{a'_s}\Delta\beta(a'_s)
-\frac{\Delta}{3!}\beta(a') \partial_{a'} \left(\Delta \beta(a')\partial_{a'}\Delta \beta(a') \right) +
\ldots
\ea
This is the way to write the  RG solution for $\bar{a}(t)$
through $\exp\left(-\Delta\, \beta(a)\partial_{a}\right)[\ldots]\mid_{a=a'}$ --- the operator of
shift (see \cite{Mikhailov:2004iq,Kataev:2014jba} and refs therein).
For the shift $\Delta$ (in the logarithmic scale) in Eq.~(\ref{eq:RGrearrange}) we set $\Delta = \Delta(a_s)$.
The latter can be expanded again in a perturbation series in terms of the rescaled charge $a'_s\beta_0$, as described in \cite{Mikhailov:2004iq},
\ba \label{Delta}
~t' &\equiv & t-\Delta, \nonumber \\
&&~~~~~ \Delta\equiv\Delta(a'_s)=\Delta_{0} + a'_s\beta_0~ \Delta_{1}
+ (a'_s\beta_0)^2~ \Delta_{2} + \ldots.
\ea
The first term $\Delta_0$ here is the standard scale shift, \eg, it may be  a part of $c_2$ that is used
in the BLM case $\Delta_0= c_2[1]$, as was discussed in the Introduction.
Of course, the $\Delta_0$ inters in higher orders of PT also, see the first rows in each cells of Table\,\ref{Tab:r_n.d_k}.
The second term of the expansion in Eq.(\ref{Delta}) gives the first correction\footnote{
The first attempt to obtain the $a_s$-correction to $\Delta_0$ was done in \cite{Grunberg:1991ac} }
$\sim \Delta_1$, \etc
Following the way we get a new parameter $a'_s$ in this expansion, which again depends on $\Delta$.
In other words, $a'_s = \bar{a}_s(t') = a_s(t - \Delta(a'_s))$, and
one should solve this equation with respect to $a'_s$.
Re-expansion of $a_s$ in terms of $a'_s$ and $\Delta_k$ leads to rearrangement of the PT series for
quantity $C(a_s)= \sum_i a^i_s c_i \to \sum_i (a'_s)^i c'_i$.
The new primed coefficients $c'_i$  depend on $\Delta$ and can be obtained via the linear transform
$c'_i= B_{ij}c_j$, where $B_{ij}\equiv \hat{B}$ is a triangular matrix, illustrated in Table\,\ref{Tab:r_n.d_k}.
In this notation, the quantity  $C$ from Eq.~(\ref{eq:RGrearrange}) (at $Q^2=\mu^2$) transforms to $C'(a'_s)$,
\begin{equation}
C(a_s) = \sum_{i\geqslant 0} a^i_s c_i \to C'(a'_s)=\sum_{i\geqslant 0} (a'_s)^i c'_i = 1+\sum_{i,j\geqslant 1} (a'_s)^i B_{ij}(\Delta) c_j \,,
\label{eq:C-pt-final}
\end{equation}
when the normalization scale $\mu$ is transformed $\mu \to \mu' $.
The elements $B_{ij}$ appear as a composition of transformations in Eq. (\ref{eq:RGrearrange}), together with the expansion of $\Delta$ in Eq. (\ref{Delta}),
and then by rearranging the power series of $a'_s$.
\begin{table}[h]
\caption{The first used
 elements of the triangular matrix $B_{ij}$,~new PT coefficients $c'_i= B_{ij}c_j$
\label{Tab:r_n.d_k}}
\begin{tabular}{|p{45mm}|p{30mm}|p{30mm}|p{30mm}|}\hline
                                   &                              &                             & \\
                      \centerline{$\bm{1}$}
                                   &  \centerline{$0$}            & \centerline{$0$}            &\centerline{$0$}
\\ \hline
                                   &                              &                             & \\
          \centerline{$-\beta_0 \Delta_{0}$}
                                   & \centerline{$\bm{1}$}
                                     & \centerline{$0$}           & \centerline{$0$}
                                     \\ \hline
            \centerline{$\Ds-\beta_1\vphantom{^{\big|}}\Delta_{0} +\beta_0^2\vphantom{^{\big|}}\Delta_{0}^2\vphantom{^{\big|}_{\big|}}\vphantom{^{\big|}_{\big|}}$}
                                  \centerline{$\Ds -\beta_0^2\vphantom{^{\big|}}\Delta_{1}\vphantom{^{\big|}}$} 
                                   &\centerline{$-2\beta_0 \Delta_{0}\vphantom{^{\big|}}$}
                                   & \centerline{$\bm{1}\vphantom{^{\big|}}$}
                                                                                &  \centerline{$0\vphantom{^{\big|}}$} 
\\ \hline
                     \centerline{$-\beta_2\Delta_{0}\vphantom{^{\big|}}-\beta_0^3\vphantom{^{\big|}}\Delta^3_{0}\vphantom{^{\big|}_{\big|}}+\frac{5}2\beta_0\beta_1\Delta^2_{0}$}
                     $+\frac{5}{2}\beta_0^3\Delta_{0}\Delta_{1}-\beta_0\beta_1\Delta_{1}\vphantom{^{\big|}}$
                     \centerline{$\Ds-\beta_0^3\Delta_{2}\vphantom{_{\big|}}$}
                                   &
                                  $\Ds-2 \beta_1\vphantom{^{\big|}}\Delta_{0}\vphantom{^{\big|}_{\big|}}$
                                  $\Ds+3 \beta_0^2\vphantom{^{\big|}}\Delta_{0}^2\vphantom{^{\big|}_{\big|}}$
                                  \centerline{$\Ds-2 \beta_0^2\vphantom{^{\big|}}\Delta_{1}\vphantom{^{\big|}_{\big|}}$}
                                                & \centerline{$-3\beta_0 \Delta_{0}\vphantom{^{\big|}}$}
                                                             &\centerline{$\bm{1}\vphantom{^{\big|}}$}
\\ \hline
\end{tabular}
\end{table}
\subsection{The perturbation series for RGI quantity under RG transformation}

Below, in the square brackets we write explicitly the elements of $c'_i=B_{ij}c_j$:
\begin{subequations}
\label{eq:rearrange}
  \ba
\!\!\!\!a_s^1\cdot c_1 \to&\!\!\!\!\!\!\!\!\!a_s'^1\cdot [c'_1=& \bm{1}]; \nonumber  \\
\!\!\!\!a_s^2\cdot c_2 \to&\!\!\!\!\!\!a_s'^2\cdot \big[c'_2(\Delta_{0})=& \bm{c_2} -1\cdot \beta_0 \Delta_{0}\big]; \label{1-stage-2} \\
\!\!\!\!a_s^3\cdot c_3 \to&a_s'^3\cdot \Big[c'_3(\Delta_{0},\Delta_{1})=&\bm{c_3} -c_2\cdot 2\beta_0\Delta_{0} -1\cdot \left(\beta_1\Delta_{0}- \beta_0^2\Delta_{0}^2+\beta_0^2\Delta_{1}\right)\bigg];
 \label{1-stage-3}\\
\!\!\!\!a_s^4\cdot c_4 \to&\!\!\!a_s'^4\cdot \Big[c'_4(\{\Delta_{i}\}_0^2)=&\bm{c_4} -
c_3\cdot 3\beta_0\Delta_{0} -c_2\cdot \left(2\beta_1\Delta_{0}-3\beta_0^2\Delta_{0}^2+2\beta_0^2\Delta_{1}\right) - \label{1-stage-4}\\
\!\!\!\!&&\!\!\!\!\!\!\!\!  1\left(\!\!\beta_2\Delta_{0}\!+\!\beta_0\beta_1\Delta_{1}
\!-\! \frac{5}2\beta_0\beta_1\Delta_{0}^2\!-\! \frac{5}{2}\beta_0^3\Delta_{0}\Delta_{1}\!+\!\bm{\beta_0^3\Delta^3_{0}}\!+\!\beta_0^3\Delta_{2}\!\right)\!\bigg]; \nonumber\\
\!\!\!\!a_s^5\cdot c_5 \to&\!\!\!a_s'^5\cdot \Big[c'_5(\{\Delta_{i}\}_0^3)=&\bm{c_5} -
c_4\cdot 4\beta_0\Delta_{0} -c_3\cdot \left(3\beta_1\Delta_{0}-6\beta_0^2\Delta_{0}^2+3\beta_0^2\Delta_{1}\right) -\label{1-stage-5} \\
\!\!\!\!&&\!\!\!\!\!\!\!\!c_2\left(\!2\beta_2\Delta_{0}\!-\!7\beta_0\beta_1\Delta_{0}^2\!+\! \ldots\!+2\beta_0^3\Delta_{2}\! \right)  -1\left(\!\beta_3\Delta_{0}\!+\ldots \!+\beta_0^4\Delta_{3}\!\right)\!\bigg];\nonumber\\
\ldots&&\ldots \nonumber \\
\!\!\!\!a_s^n\cdot c_n \to&a_s'^n \Big[c'_n(\{\Delta_{i}\}_0^{n-2})\!\!=&\!\!\bm{c_n} -
c_{n-1}\cdot (n-1)\beta_0\Delta_{0} -\ldots \mp(\beta_0\Delta_{0})^{(n-1)}-  \nonumber \\
   &&\phantom{\bm{c_n} - c_{n-1}\cdot (n-1)\beta_0\Delta_{0} -\ldots \mp(\beta_0\Delta_{0})} \beta_0^{(n-1)}\Delta_{n-2}\Big]\,.
\ea
 \end{subequations}
We extend the matrix $B_{ij}$ a bit further in the next order (\ref{1-stage-5}) for illustration here.
It should be noted that the term $\bm{\beta_0^3 \Delta^3_0}$ in Eq.(\ref{1-stage-4}) was erroneously omitted in Eq.(7) of \cite{Kotlorz:2018bxp}.

The limit $\Delta = \Delta_0$, $\Delta_{k\geqslant 1}=0$ brings us back to the standard form of transformation $\mu^2 \rightarrow \mu'^2$
in the logarithmic notation  $\Delta_0$.
At the same time radiative corrections to $\Delta$ in Eq. (\ref{Delta}), give us the $n - 2$ new parameters $\{ \Delta_i \}_0^{n - 2} \to \Delta$ in each order $n$.
Therefore, every coefficient $c_n'$ acquires a new parameter $\Delta_{n-2}$.
The price of these opportunities is the need to solve the equation $a'_s = a_s (t - \Delta(a'_s))$ for each value of
the scale parameter $t$  numerically.
It seems tempting to think that the parameters $\{ \Delta_i \}_0^{n-2}$ completely determine the values of coefficients $\{ c_2' - c_n' \}$.
We can now adjust the parameters $\Delta_0, \Delta_1, \Delta_2, \ldots$ of the new scale $\mu'$, to fit the coefficients $c_2', c_3', c_4', \ldots$ according to a chosen goal of optimization.
In the next section we will discuss constrains on the $\{\Delta_{i}\}$.

\section{Perturbation hierarchy as a filter for $\{ \Delta \}$ parameters}
\label{sec:sec3}

Let us apply the general scheme developed in the previous section to the relevant quantity
$C$ starting from the appropriate conditions for the truncated PT series in Eqs.~(\ref{Delta}-\ref{eq:C-pt-final}).
At first sight, it might seem that one can choose any value for the new scale $\mu'$ and, therefore,
the parameters $\{\Delta_{i} \}$ in Eq.~(\ref{Delta})
might look like not constrained from any side, but that is an illusion.
We will formulate the necessary conditions for the PT expansions below.
In order to satisfy the reliability requirements for the PT expansion, we demand a natural hierarchy,
inequalities (A -- C) for its successive terms:

\begin{itemize}
\item[\textbf{(A)}]
For the terms of PT expansion of $\Delta(t)=\Delta_{0} + a'_s\beta_0\Delta_{1}
+ (a'_s\beta_0)^2\Delta_{2} + \ldots$ in Eq.~(\ref{Delta}) there should be
\be \label{eq:Delta-cond}
|\Delta_{0}|> |A'\Delta_{1}|> |A'^2 \Delta_{2}|> |A'^3 \Delta_{3}|> \ldots,\;\;\;\;\;
 \text{where}\;\; A'\equiv\beta_0 a'_s\;,
\ee
which means that the next term of this PT expansion cannot be larger than the previous one.
These inequalities assume nonlinear conditions for $\Delta_i$,
which become more restrictive for the case of $\Delta > 0$ due to asymptotic freedom,
$A' \simeq 1/t'$, where $t'=t-\Delta$
is defined in Eq. (\ref{Delta}).
Conversely, for $\Delta < 0$, the domains can extend indefinitely to the left.
\item[\textbf{(B)}]
For PT expansion in Eq.~(\ref{eq:C-pt-final}) we impose conditions with respect to $c'_i$,
which are similar to Eq.~(\ref{eq:Delta-cond}):
 \be
 \frac{1}{|c^\text{ns}_1|} = \frac{1}{4}> a'_s > a'^2_s \bigg{|}c'_2\bigg{|} > a'^3_s \bigg{|} c'_3\bigg{|}
> a'^4_s \bigg{|}c'_4\bigg{|}> \ldots \, \label{eq:4.1a-cond}
\ee

 where one can add to admissible regions a special limit of (\ref{eq:4.1a-cond}),
 \be
   c'_2= c'_3=c'_4 = 0\,,
 \ee
  that can be reached at least asymptotically.
\end{itemize}
We acknowledge at the same time that it is possible to propose and justify more restrictive conditions than those that are more or less obvious in (\ref{eq:Delta-cond}) and (\ref{eq:4.1a-cond}).
We emphasize that some constraining conditions on the PT series in (\ref{Delta}) and (\ref{eq:rearrange}) are required in any case.
The new coefficients $c'_i$ are given by Eq.~(\ref{eq:rearrange}),
while the explicit forms of the initial coefficients $c_{i}$ are presented in Appendix \ref{App:A}.
The running $\bar{a}_s$ has an asymptotic expansion, Eq.~(\ref{eq:beta.new.4L}) of Appendix \ref{App:C}, or
can be taken from the numerical solution of Eq.~(\ref{eq:beta3.new}).\\
\textbf{(C)} To fix the PT domain of applicability, we put for the logarithmic variable $t'=t-\Delta(t')$
the appropriate lower bound of $\mu'^2$ at $\mu'^2=\mu_0^2$. We take the conventional $\mu_0^2 = 1$ GeV$^2$ that corresponds to
$t_{\mu_0}=\ln\left(\mu_0^2/\Lambda_{qcd}^2\right)\approx 2.291$ at $\Lambda_{qcd}=\Lambda_{(4)}=0.318$\,GeV;
therefore
\ba \label{eq:PTbound}
t, t'\geqslant t_{\mu_0}&&  \Rightarrow t-t_{\mu_0} \geqslant \Delta(t')=\Delta_{0} + A'\cdot \Delta_{1} + A'^2\cdot \Delta_{2}\,.
\ea
We assume Eqs.~(\ref{eq:Delta-cond}, \ref{eq:4.1a-cond}, \ref{eq:PTbound}) to be \textit{necessary conditions}.
Next, we will scan $t$ in the practically interesting interval $2.3 < t \leqslant  11.32$
($1 < \mu^2 \leqslant 2227,\, M^2_Z=8316$~GeV$^2$) that covers the scales from $m^2_\tau, t_\tau \approx 3.44$ to $M^2_Z, t_{Z}\approx 11.32$.
Very large scales $\mu'^2$ corresponding to the large values of $-\Delta$ contradict the PT hierarchy of the terms in (\ref{eq:4.1a-cond}) and are therefore forbidden.
The latter can be easily seen from Eq.(\ref{eq:rearrange}) if we consider it for large  values of $-\Delta_0=\ln(\mu'^2/Q^2)$.
In this case, $a'^n_s c'_n \sim   a'^n_s \ln^{n-1}\left(\mu'^2/Q^2 \right) \sim 1/\ln(\mu'^2/\Lambda^2_{qcd})$
and the PT hierarchy is lost because the contributions of all PT terms become asymptotically equal as $\mu'^2/\Lambda^2_{qcd} \to \infty$.
We will localize the domain of the parameters $\{\Delta_{0},\Delta_{1},\Delta_{2}\}$,
where the constraints in Eqs. (\ref{eq:Delta-cond}), (\ref{eq:4.1a-cond}),
and (\ref{eq:PTbound}) are simultaneously satisfied.
These conditions form the admissible domain in the $\{\bm{\Delta}\}$-space at every value of $t$ (marked with its own color),
see colored lines/strips/domains in Figs.\ref{fig:bsr1}, \ref{fig:bsr-minmax7}.
Here we will explore the size of $\Delta_0$ in relation to $t$, $|\Delta_0| \lesssim t$.
More precisely, our scale $\mu'$ is constrained by $\mu'^2 \lesssim M_Z^2 = \text{max}(Q^2, s)$,
which means that we use $|\Delta| < t_Z$\footnote{
We have imposed more restrictive conditions on $\Delta$ in \cite{Kotlorz:2018bxp}, namely, $|\Delta| \leqslant t$,
which led to a special limitation of the size of the admissible domains.} for all $\Delta$.

\section{On optimization of perturbative series in example of $C_\text{Bjp}$}
\label{sec:sec4}
\begin{figure}[ht]
  \includegraphics[width=0.53\textwidth]{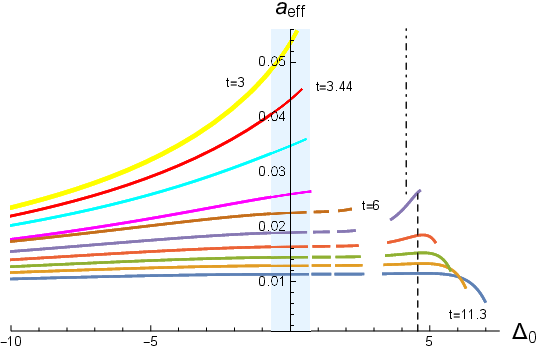} 
   \includegraphics[width=0.46\textwidth]{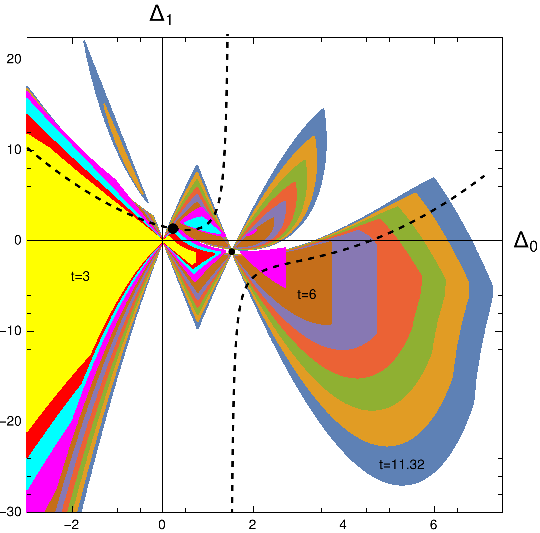}                              
    \vspace*{-4mm}
       \caption{\label{fig:bsr1} \footnotesize
   The admissible domains of $\{\Delta \}$ for $C_\text{Bjp}$ at each value of $t$ are shown by the lines or
surfaces of a specific color,  which are bounded on the right and extend unbounded to the left.
   Left: 1D case. The $a_\text{eff}$ corresponds to the effective charge in Eq.(\ref{eq5.2}), \ie, the sum of all radiative correction equal to $-4\, a_\text{eff}$.
 The dashed (dashed-dotted) black lines  correspond the condition Eq.(\ref{eq:c4=0c}):\, $n_f=5\,(4),\Delta_0=4.57\, (4.16)$;
 the blue vertical strip -- $|\Delta_0| \leqslant \ln(2)$.
 Right: 2D domains at different $t$ (having different colors) increase in the size with $t$  to the right (left).
  $\bullet$ -- limit point $c'_2\!=\!c'_3\!=\!c'_4\!=\!0$ at $\Delta_0\!=\!1.52, \Delta_1\!=\!-1.18, \Delta_2\!=\!1.77$ in 3D.
  The dashed black lines correspond to condition in Eq.(\ref{eq:delta0-1}), the black disk \ding{108} -- the conditions $c'_3\!=\!c'_4\!=\!0$ in Eq.
  (\ref{eq: 11d}) at $n_f=5$.}
\end{figure}

Here we will explore a few examples of possible ``optimization'' that depend on the coefficients $c_i$ as a whole,
 rather than the knowledge of the details of their $\{\beta \}$-expansion.
In this way, we test the limits and possibilities for modifying PT within the  aforesaid general conditions
\textbf{(A-C)}.
We would like to preliminary note here two evident properties of $\Delta(t)$ as $t$ increases, which
are illustrated in Fig.\,\ref{fig:bsr1}(Left part -- the dimension $D=1$, Right -- $D=2$), at $t=7$ see Fig.\ref{fig:bsr-minmax7},
and Fig.\ref{fig:bsr-3d} ($D=3$): \\
(1) The admissible domain  corresponding to a certain $t$ also increases with $t$.
Smaller domain are superimposed on the next larger ones in order of increasing $t$.
The expansion of the domains is the final manifestation of asymptotic freedom; \\
(2) The dependence of the sum of radiative corrections on the value of $\Delta$ significantly decreases as $t$ increases.

The accepted conditions \textbf{(A)}, \textbf{(B)}, and \textbf{(C)} from Sec.\ref{sec:sec3} shape ``islands'', ``peninsulas'' and ``channels'' that
coordinately appear in all dimensions of the domains in $\Delta$ spaces, as it is shown in Figs.\ref{fig:bsr1},\ref{fig:bsr-minmax7},\ref{fig:bsr-3d}.
Below we show a useful correspondence between the $Q^2$ and $t$ scales starting with $Q^2=2$\,GeV$^2, t=3$: \\
~~~~$t~~=~3,~~~~~~~3.44,~~~~~~~4,~~~5,~~~6,~~~~~~7,~~~~~8,~~~~ 9,~~~~ 10,~~~~~~ 11.32$\,: \\
 $Q^2=2,~ (m_\tau\!=\!1.77)^2,~5.5,~15,~40.8,~ 110.8,~301,~819, ~2227, (M_Z=91.19)^2$ GeV$^2$.\\
 \vspace{-4mm}

\subsection{The constraint on the highest expansion coefficients $c'_n =0$ }
We start by considering the possibility of removing the last known perturbative term, $c'_n$, of the expansion by selecting an appropriate $\mu'$ in the notation of the parameters $\Delta$, see Eq.(\ref{Delta}).
So one can impose this condition on the last known coefficient, $c'_4=0$, following transformation in Eq.(\ref{1-stage-4}),

\ba
\!\!\!\!\!\!\!\!c'_4&\!=\!&c_4- \Delta_{0} \left( 3\,c_3\beta_0 +2 c_2\, \beta_1+\beta_2\right) +\nonumber \\
\!\!\!\!\!\!\!\!&&
\Delta_{0}^2\! \left(3 c_2\beta_0^2 + \frac{5}2\beta_0\beta_1\right)-
\Delta_{1} \left( c_2\, \beta_0^2+\beta_0\beta_1\right)+
 \Delta_{0}\Delta_{1} \frac{5}2\beta_0^3\!-\!\Delta_{2}\,\beta_0^3\!-\!\Delta^3_{0}\,\beta_0^3=0\,. \label{eq:delta0-1-2}
\ea
This condition provides  variety of constrains on the independent parameters $\{\Delta_{0},\Delta_{1},\Delta_{2}\}$,
\begin{subequations}
\label{eq:c4=0}
\ba
&&c'_4=0: \label{eq:c4=0a}\\
D=1,&\Delta_{1}=\Delta_{2}=0 &c'_4=c_4 -
\Delta_{0} \left( 3\,c_3\beta_0 +2 c_2\, \beta_1+\beta_2\right) +
\Delta_{0}^2 \left(3 c_2\beta_0^2 + \frac{5}2\beta_0\beta_1\right) \nonumber\\
&& \phantom{c'_4=c_4}-\Delta^3_{0}\,\beta_0^3=0 ; \label{eq:c4=0b}\\
&&n_f=4, \Delta^{(4)}_0=2.56;~n_f=5, \Delta^{(5)}_0=3.46; \label{eq:c4=0c} \\
D=2,&\Delta_{2}=0 &c'_4=c_4 -
\Delta_{0} \left( 3\,c_3\beta_0 +2 c_2\, \beta_1+\beta_2\right) +
\Delta_{0}^2 \left(3 c_2 \beta_0^2 + \frac{5}2\beta_0\beta_1\right)+ \nonumber \\
&&\phantom{c'_4=c_4}-\Delta_{1} \left( c_2\, 2\beta_0^2+\beta_0\beta_1\right)+
 \Delta_{0}\Delta_{1}\frac{5}{2} \beta_0^3-\!\Delta^3_{0}\,\beta_0^3=0 \,. \label{eq:delta0-1}
\ea
 \end{subequations}

\begin{figure}[ht]
\includegraphics[width=0.46\textwidth]{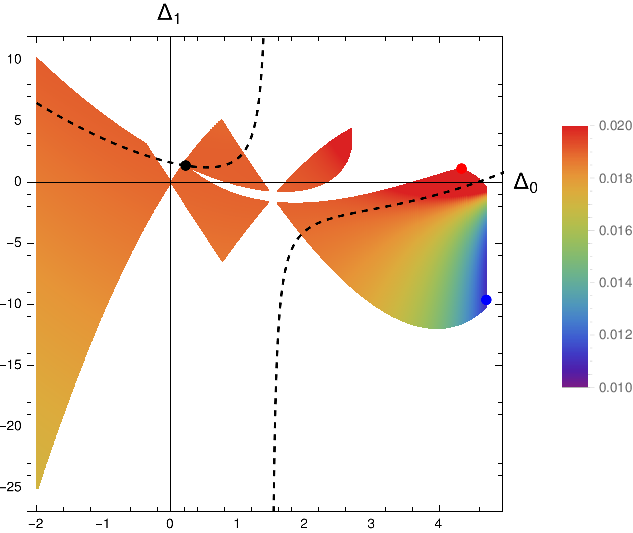} 
\includegraphics[width=0.53\textwidth]{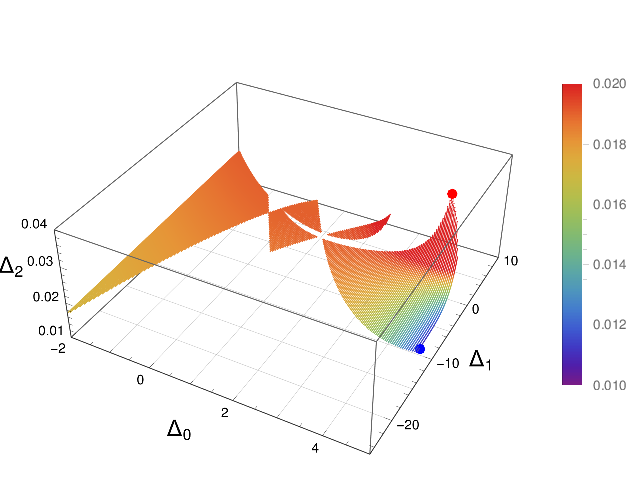} 
\vspace*{-5mm}
  \caption{ \label{fig:bsr-minmax7} \footnotesize The admissible domain of $\{\Delta\}$ for $C_\text{Bjp}$ at $t=7$ (violet domain in
  Fig.\ref{fig:bsr1}).
  The color of the domain here means the value of $\alpha_\text{eff}$,
  the cooler the color, the smaller  $\alpha_\text{eff}$.
  The red-$\red{\bullet}$/blue-$\blue{\bullet}$ points are max/min of $\alpha_\text{eff}$ in the domain.
  Left: 2D domain is crossed by the dashed  lines of  the condition in (\ref{eq:delta0-1-2}),
  the black disc $\bullet$ corresponds to (\ref{eq: 11d}), \ie, $c'_3=c'_4=0$.
  Right: the same 2D domain is nested into 3D box where the $\alpha_\text{eff}$
  values are shown along the axis $z$.
}
 \end{figure}

  As it is demonstrated for $C_\text{Bjp}$ in Figs.\ref{fig:bsr1}, \ref{fig:bsr-minmax7}
  the conditions (\ref{eq:c4=0c},\ref{eq:delta0-1}) are fulfilled  practically for  values of $t\geqslant 7$,
  see the black dashed lines there.
  For the simplest $1D$ case there are crossing points of these (vertical) lines and the admissible domains starting with
  $t\geqslant 7$,
  see Fig.\ref{fig:bsr1} Left. This cancel of $c_4$ is not very helpful  because the corresponding $t'$ becomes risky close to the
  bound of PT $t_{\mu_0}$ at $a_s(t'=t-\Delta_0^{(5)})$.

\begin{figure}[ht]
\includegraphics[width=0.7\textwidth]{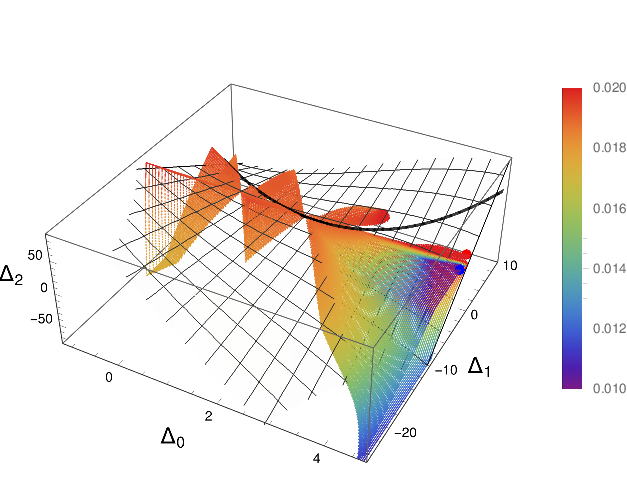} 
\vspace*{-5mm}
  \caption{ \label{fig:bsr-3d} \footnotesize The 3D admissible domain of $\{\Delta\}$ for $C_\text{Bjp}$ at $t=7$.
  The color  here means the value of $\alpha_\text{eff}$,
  the cooler the color, the smaller  $\alpha_\text{eff}$,
  as well as for the 2D domain in Fig.\ref{fig:bsr-minmax7}.
The 3D domain is crossed by the net of the condition $c'_4=0$,
  while the black curve on the net corresponds to the condition $c'_3=c'_4=0$.
  The red-$\red{\bullet}$/blue-$\blue{\bullet}$ points are max/min of $\alpha_\text{eff}$ in the domain.
}
 \end{figure}

 Unlike the previous,
 the contribution from  $c_4$  can be eliminated by rather moderate values of shift $\{\Delta_0, \Delta_1\}$ of the argument $t$ in $2D$ case,
 see Fig.\ref{fig:bsr1}Right.
 Indeed,  one can get rid of $c_4$ by  moving along the dashed black line (its left branch)
  even within $|\Delta_0| \lesssim 1$ for all $t$.
 A similar situation occurs for the 3D case of the parameters   $\{\Delta_0, \Delta_1, \Delta_2\}$, see Fig.\ref{fig:bsr-3d}.
  For the latter case at $t=7$, the gray net corresponding to the condition in Eq.(\ref{eq:delta0-1-2}) crosses the body of this domain well.

Take another requirement: two consecutive terms of PT, $c'_3$ and $c'_4$, are equal to zero, which provides the relations for $\Delta_0$, $\Delta_1$, and $\Delta_2$,
\begin{subequations}
 \label{eq:c4=c3=0}
 \ba
c'_3&=&c'_4\!=\!0: \label{eq: 11a}\\
c'_3&=&c_3 -\Delta_{0}\beta_0\left( 2 c_2+ \frac{\beta_1}{\beta_0}\right) + \beta_0^2\Delta_{0}^2-\beta_0^2\Delta_{1}\!=\!\!0;\label{eq: 11b} \\
\!\!\!\!\!c'_4\!&=&\!c_4\!-\! \Delta_{0}\beta_0\! \left(\!3 c_3\! +\!2 c_2 \frac{\beta_1}{\beta_0}+\frac{\beta_2}{\beta_0}\right)\!+\!
(\Delta_{0}\beta_0)^2\!
\left(\!3c_2\!+ \frac{5}2\frac{\beta_1}{\beta_0}\right)\!-\!\Delta_{1}\beta_0^2
\bigg[\! 2 c_2+\frac{\beta_1}{\beta_0} - \frac{5}{2} \Delta_{0}\beta_0 \bigg]\! \nonumber\\
&&-\! (\Delta_{0}\beta_0)^3\! -\!\beta_0^3\Delta_{2}\!=\!0. \label{eq: 11c}
\ea
 The  condition Eq.(\ref{eq: 11a}) can be realized even at $\Delta_2=0$ for a couple of points that are presented in Eq.(\ref{eq: 11d}),
\ba
\text{for}\,\Delta_2=0,\,&&\text{at}\,  \bigg\{
  \begin{array}{l}
  n_f=4, \Delta^{(4)}_0=0.103; \Delta^{(4)}_1=2.803; \label{eq:4.3d} \\
  n_f=5, \Delta^{(5)}_0= 0.220, \Delta^{(5)}_1= 1.368\,,
\end{array}
\label{eq: 11d}
\ea
 \end{subequations}
 see black point (at $n_f=5$) in Figs.\ref{fig:bsr1}Right and \ref{fig:bsr-minmax7}Left.
Therefore, for a wide range of $t$ (at least for $t \geqslant 4$), we can apply the conditions $c'_3 = c'_4 = 0$,
rearranging their contributions in a new coupling constant $a_s'$  as follows:
\begin{subequations}
 \label{eq:conditions_c3=c4}
 \ba
 C\left(t,\{\Delta \}\right)&=&
1+c_1^\text{ns}\Big(1\,a'_s+c'_2(\Delta_0 )\,a_s'^2 + 0 + 0\Big), \label{eq:4.4a}\\
 \text{where,}&& c_2'(\Delta_0 )= c_2-\beta_0\cdot \Delta^{(4,5)}_0 \stackrel{n_f=5}{\longrightarrow}\left(\frac{35}3 \right)-\beta_0\cdot 0.22
 ;        \label{eq:4.4b}   \\
&& a_s' = \bar{a}_s(t')=\bar{a}_s(t - \Delta(a'_s));~ \Delta(a'_s)=\Delta^{(4,5)}_0+a_s'\Delta^{(4,5)}_1\,. \label{eq:a'_s}
  \ea
   \end{subequations}
The values of the $\Delta_i$ parameters in Eq.(\ref{eq:4.4b},\ref{eq:a'_s}) are taken from Eq.(\ref{eq:4.3d}) (at $n_f=5$); moreover,
 at this condition $c'_3=c'_4=0$ we also find that new $c'_2 < c_2$. The behavior of $a'_s(t)$ in Eq.(\ref{eq:a'_s}) is illustrated in Fig.\ref{fig:alpha-mod} in  Appendix\,\ref{App:C}. Results  similar to those above  are obtained for RGI $D_\text{NS}$ at the conditions on the coefficients $d_3=d_4=0$ and are presented in Appendix\,\ref{App:B}, Eq.(\ref{eq:d3=d4=0}).
 The 3D curve that is determined by the relations in Eqs.(\ref{eq: 11b}, \ref{eq: 11c}) penetrates through part of the bodies of the admissible domains in $\{\bm{\Delta} \}$ starting with $t\geqslant 6$, see black curve in Fig.\ref{fig:bsr-3d}.
The price for the series ``improvement'' in (\ref{eq:4.4a}) is the need to solve the equation $a'_s\!=\!a_s (t - \Delta(a'_s))$
in (\ref{eq:a'_s}).

 Finally, we conclude that the convergence of the PT series for $C_\text{Bjp}$ can be improved by fulfilling the conditions $c'_4=0$ in (\ref{eq:c4=0}) or $c'_3 = c'_4 = 0$ in (\ref{eq:4.3d},\ref{eq:conditions_c3=c4}), when choosing the appropriate shifts $\{\Delta^{(4,5)}_0, \Delta^{(4,5)}_1\}$.
Similar results to Eq.(\ref{eq:conditions_c3=c4}) are obtained for the PT series of $D_\text{NS}$, as shown in Eq.(\ref{eq:d3=d4=0}).
Such new perturbation series for $C_\text{Bjp}$ and $D_\text{NS}$ can be further applied to the processing  the corresponding experimental data.
\subsection{Extrema of the sum of radiative corrections}
 \label{sec:6b}
Let us now investigate  the minimum/maximum radiative corrections to $C$ by varying the scale $\mu'$
within the admissible domain of the parameters $\{\bm{\Delta}\}$.
In other words, we want to get the result closer to the ``partonic" picture, or, on the contrary,
be as far away from it as possible due to radiative corrections.
We numerically find the extrema of the function $a_\text{eff}/\pi$, which
accumulates all of the known radiative corrections up to order $\alpha_s^4$,
\begin{subequations}
 \ba
 C(t', a'_s)&=&1+c_1\frac{\alpha_\text{eff}(t;\{\bm{\Delta} \})}{\pi} \,, \label{eq5.1} \\
&&\frac{\alpha_\text{eff}}{\pi}=
a_s' \left(1+a'_s c'_2+ (a'_s)^2 c'_3 + (a'_s)^3  c'_4 \right)\,. \label{eq5.2}
\ea
 \end{subequations}
The quantity $\alpha_\text{eff}/\pi$
is the auxiliary ``effective charge'' proposed in \cite{Deur:2005cf,Deur:2016cxb}.
The arguments $\{\bm{\Delta}\}$ are taken within the admissible domain corresponding to a certain  $t$.

It is instructive to start the discussion with the popular test for the uncertainty in the results by varying
the scale $\mu^2=Q^2$ within a range $\Delta \mu^2 = (Q^2 / 2, 2 Q^2)$ around $Q^2$.
This test is based on the hope that the crucial scale lies within this  range.
The range corresponds to the interval $|\Delta_0| \leqslant \ln(2) \equiv \Delta_0^{(2)}$,
which serves as a universal ``scale'' that, however, is arbitrary and has no relation to the proper
characteristics of the RGI-quantity under consideration.
This test shows visible variation for relatively low $t \leqslant 5$ despite the fact that the value of $\Delta_0^{(2)}$ is less than,
\eg, the BLM scale $\Delta_0^\text{BLM}=c_2[1] = 2$, which may be crucial and is specific to the $C_{\text{Bjp}}$
(see the discussion about the ``true scale'' in Sec.\ref{sec:intro}).
It looks natural that such  uncertainties become smaller with growing $t$: really,
the results $\alpha_\text{eff}(t)$ do not change visibly within the strip $|\Delta_0| \leqslant \ln(2)$ for $t \geqslant 7$.

An interesting example is provided by the behavior of  $\alpha_\text{eff}$ starting with $t = 7$,
see Fig.\ref{fig:bsr1} Left ($D=1$),
which shows anomalous increases up to $0.0265 (+40\%)$  vs the value of $0.019$ at $\Delta_0 = 0$,
when  $\Delta_0$ approaches  the low boundary $t-t_{\mu_0}$ ($t_{\mu_0}\simeq 2.3$) of this admissible domain.
The position of this max happens to be close to the condition $c'_4=0$ at $\Delta_0=4.57$.
For the 2D parameterization the minimum of $\alpha_\text{eff}$  reaches $0.011 (-42\%)$ --- the blue point on the boundary of
the domain in Fig.\ref{fig:bsr-minmax7},
while the maximum reaches $0.038 (+100\%)$ --- the red point is also near the boundary here.
In Fig.\ref{fig:bsr-minmax7} Right, the same 2D domain is embedded in the 3D box,
where the $\alpha_\text{eff}$ values are shown on the $z$-axis for illustration.
The max value of $\alpha_\text{eff}$ occurs at  $t''$, see Eq. (\ref{eq:minmaxarg}),
 that is very close to the lower boundary of applicability,
\begin{subequations}
 \ba
\!\!\!\!\!\!\!\!\!\!\!\!\!\!\bm{\downarrow}\text{way to the min};~~ 0.011 &<\alpha_\text{eff}(t=7)\approx 0.0190 < & 0.038;~ \text{way}\bm{\downarrow} \text{to the max}, \label{eq:minmaxpoints}\\
\!\!\!\!\!\!\!\Delta'\to\{4.7,\, -9.62\}~\bm{\downarrow}~&  (\Delta_0=0,\Delta_1= 0) &~\bm{\downarrow} \Delta''\to\{4.33,\, 1.12\} \label{eq:2Dways} \\
 \bar{a}_s(t'=4.128)=0.0248          &                                       &\bar{a}_s(t''=2.3)=0.0431\,; \label{eq:minmaxarg}
\ea
 \end{subequations}
 therefore, this estimate is not reliable.

The picture similar to the $2D$ case is repeated  for the $3D$-parameterization presented in Fig.\ref{fig:bsr-3d}.
The extrema: max -- the red point $\red{\bullet}$, and min -- the blue one $\blue{\bullet}$, occur approximately at the
close points  of the argument $t'$, $t''$ (recall that $t'(t'') = t - \Delta' (\Delta'')$) which tend to the low limit of acceptability $t_{\mu_0}\approx 2.3$.
Due to the latter, they are not reliable and should be considered as upper and down limits of $a_\text{eff}$ with great caution:
\begin{subequations}
 \label{eq:3Dminmax}
 \ba
\!\!\!\!\!\!\!\!\!\!\!\!\!\!\!\!\!\bm{\downarrow}\text{way to the min}; 10^{-5} &<\alpha_\text{eff}(t=7)\approx 0.0190 < & 0.0428;\, \text{way} \!\bm{\downarrow}\text{to the max}, \\
\!\!\!\Delta'\to\{4.69,~- 2.56,~7.11\} \bm{\downarrow}&  &\!\!\!\!\bm{\downarrow}\Delta''\to\{4.57, 1.42,\! -3.27\}  \label{eq:3Dways} \\
\!\!\!  \bar{a}_s(t'=2.41)=0.041          &                                       &\bar{a}_s(t''=2.32)=0.0428\,. \label{eq:3Dpoints}
\ea
 \end{subequations}
However, these extrema are reached in different ways in 3D space $\Delta \to\{\Delta_0$, $\Delta_1$, $\Delta_2\}$.

\section{Different types of $\{\beta\}$-expansions against the filtering background}
 \label{sec:sec5}
Here, we will review different types of $\{\beta \}$-expansions that have
been presented in the literature \cite{Mikhailov:2004iq, Mojaza:2012mf, Ma:2015dxa, Cvetic:2016rot},
as well as the related types of ``optimizations'' discussed in
\cite{Kataev:2014jba, Wu:2013ei, Mikhailov:2024mrs}, against the background
of the universal filters for $\{\Delta_i\}$ that were discussed in
Sec.~\ref{sec:sec3}.
\subsection{What does PMC mean?}
 \label{sec:sec5a}
\begin{figure}[hb]
\includegraphics[width=0.55\textwidth]{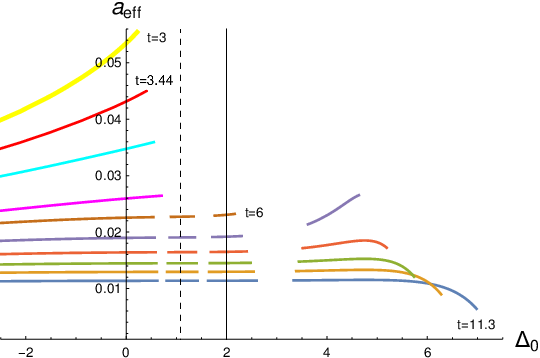}
\includegraphics[width=0.44\textwidth]{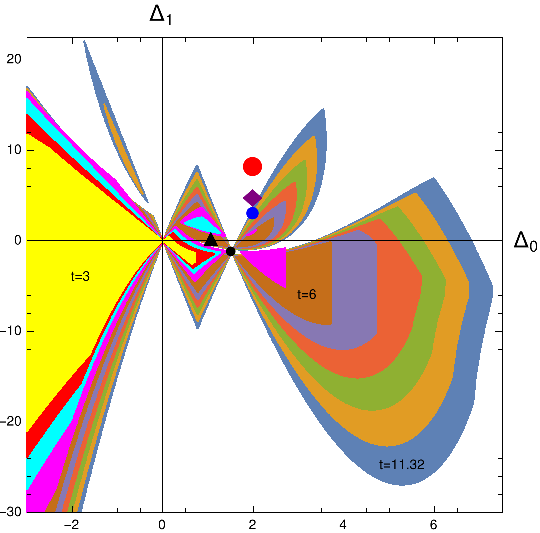} 
\vspace*{-4mm}
  \caption{ \label{fig:bsr-points} \footnotesize
  The admissible domains of $\Delta$ for the quantity $C_\text{Bjp}$ under
different series transformations.
Left: 1D case. The vertical dashed line denotes PMC \cite{Mojaza:2012mf, Ma:2015dxa},
$\Delta_0 = c_2[1]-\frac{11}{12}\approx 1.08$;
the vertical solid line corresponds to BLM and gNNA \cite{Mikhailov:2004iq},
$\Delta_0 = c_2[1] = 2$ for this order.
Right: 2D case. $\bullet$ is the asymptotic limit point $c_2=c_3=c_4=0$.
Different versions of PMC, respectively:
\black{ \ding{115} } \cite{Ma:2015dxa},
$(n_f=5(4))$, $\Delta_0 = 1.08$ and $\Delta_1\approx 0.29\, (0.58)$,
Eq.(\ref{eq:Delta});
\red{ \ding{108} } \cite{Baikov:2022zvq}, $(n_f\!=\!5)$,
$\Delta_0 = 2$ and $\Delta_1 = 8.27$; 
\amethyst{\ding{117}} \cite{Goriachuk:2021ayq,Kataev:2022iqf},
$(n_f = 5)$, $\Delta_0 = 2$ and  $\Delta_1 = 4.83$.
gNNA: \BluTn{\ding{108}} \cite{Mikhailov:2024mrs}, $(n_f = 5(4))$,
$\Delta_0 = 2$ and $\Delta_1 = 3.058\, (2.75)$ 
}
\end{figure}

The PMC proposal, suggested in \cite{Mojaza:2012mf}, uses only the single element of any
$\{\beta \}$-expansion in 
order $i$, namely, the ``scale-invariant''
part $c_i[0]$, see Appendix \ref{App:A}.
So PMC requires $c'_i=c_i[0]$, while the corresponding $\{ \Delta_j \}$ can be
sequentially extracted from the set of Eqs.(\ref{eq:rearrange}).
It should be noted that the PMC in (\ref{eq:CBJ11}) has no justification from
the perspective of optimizing the PT series.
Following the PMC, the series for the quantity $C'$ reads:
\begin{subequations}
\ba \label{eq:CBJ11}
C' &=& 1 + c_1^\text{ns} \Big(a'_s + (a'_s)^2 c_2[0] + (a'_s)^3 c_3[0] + (a'_s)^4 c_4[0]+\ldots \Big)\,; \\
&&c_1^\text{ns}=-3C_\text{F}=-4\,,
\ea
\end{subequations}
where $a'_s=a_s\left(\mu^2_\text{PMC}\right)$ and the new scale
$\mu^2_\text{PMC}$ is determined using the parameters $\Delta_{i}$.
These $\Delta_{i}$ are, in turn, determined by the remainder of $c_i - c_i[0]$
according to the set of Eqs.(\ref{eq:rearrange}).
\begin{subequations}
\label{eq:Delta}
\ba
\beta_0\Delta_{0}&=&c_2- c_2[0]=\beta_0 c_2[1], \Delta_{0}=c_2[1];\\
\beta_0^2\Delta_{1}&=&c_3- c_3[0] -\beta_1\Delta_{0}- 2\beta_0 c_2\Delta_{0}+\beta_0^2\Delta_{0}^2 \\
\beta_0^3\Delta_{2}&=&c_4- c_4[0] -
 3 c_3\beta_0 \Delta_{0}
 -c_2\cdot \left(2 \beta_1\Delta_{0}-3\beta_0^2\Delta_{0}^2+2\beta_0^2\Delta_{1}\right)\nonumber\\
&& -\left(\beta_2\Delta_{0}+\beta_1\beta_0\Delta_{1}
-\frac{5}2\beta_1\beta_0\Delta_{0}^2- \frac{5}{2}\beta_0^3\Delta_{0}\Delta_{1}\right)
\ea
\end{subequations}
All required elements to check for inclusion in the hierarchy filter are
presented in Appendices \ref{App:A} and \ref{App:B}.
The different types of the $\{\beta\}$-expansion lead to different
positions for the points/lines of $\{\Delta_0, \Delta_1 \}$, as shown for
BSR in Fig.\ref{fig:bsr-points}.
For the 1D case with $\Delta_0$ in the left panel of Fig.\ref{fig:bsr-points},
all PMC versions are admissible at $t\geqslant 6$ (see the vertical lines).
In the 2D case, $\{\Delta_0, \Delta_1\}$, shown in the right panel of
Fig.\ref{fig:bsr-points}, the symbol \black{\ding{115}} corresponds to PMC in
\cite{Ma:2015dxa}.
It is seen that at $t\geqslant 4$ this kind of PMC belongs
to the admissible domain and can be applied.
Despite this, we disagree with the PMC approach presented in
\cite{Mojaza:2012mf, Ma:2015dxa} and explicitly applied to $C_\text{Bjp}$
 in \cite{Deur:2017cvd}.
Indeed, the key elements $c_i[0]$ in Eq. (\ref{eq:Delta})
were derived from the reduced CBK relation \cite{Deur:2017cvd},
taking into account the elements $d_j[0]$ of $D_\text{NS}$.
However, this seems unconvincing due to the utilized $d_j[0]$ construction,
which is discussed below.
Moreover, the $\{\beta\}$-expansion used by the authors contradicts the
direct calculation, as discussed in \cite{Mikhailov:2024mrs}.
Therefore, we consider the result \black{\ding{115}} to be accidental.
In addition, the authors of \cite{DiGiustino:2023jiq} claimed that the PMC
procedure gets rid of the renormalon divergence.
We collect the terms from the renormalon chain into the shift
$\Delta_\text{ren}$,
\be \label{eq: renormalon}
   \Delta \to \Delta_{ren} = c_2[1] + \sum_{i=1}^{n} (a'_s \beta_0)^i c_{i+2}[i+1]\,,
\ee
which diverges factorially with $n$ as $c_{n+2}[n+1] \sim n!$
\cite{Broadhurst:1993ru}.
So the renormalon growth is simply  transferred from the coefficient function $C$ to the value of the shift
of the scale $\Delta =-\ln\left((\mu')^2/\mu^2\right)$ by means of Eq.(\ref{eq: renormalon}).
Other PMC results, \red{\ding{108}} and \amethyst{\ding{117}}, extracted from
\cite{Baikov:2022zvq} and \cite{Kataev:2022iqf} respectively fall outside the acceptable
domains, although they were not intended for optimization purposes
in the spirit of PMC.

\begin{figure}[hb]
\hspace{-5mm}\includegraphics[width=0.53\textwidth]{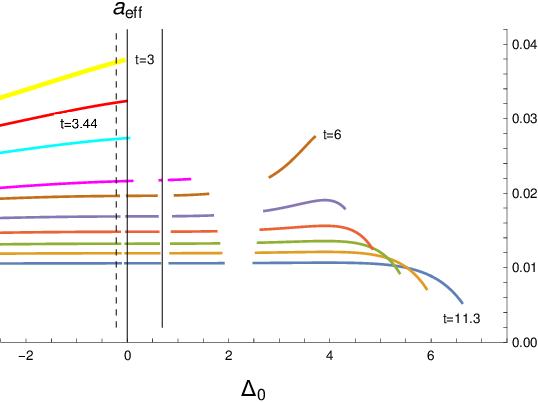}
\includegraphics[width=0.46\textwidth]{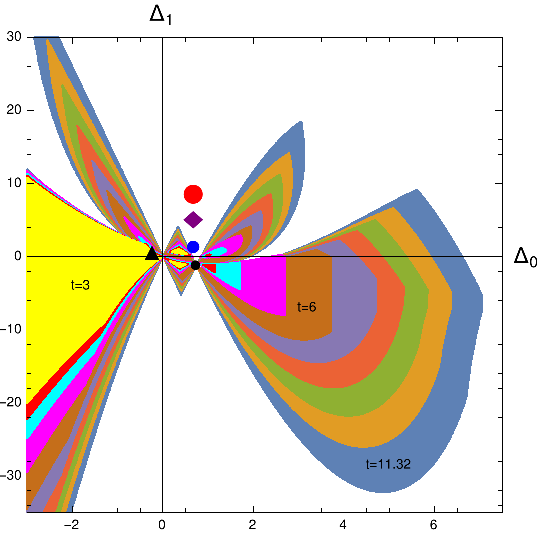} 
\vspace*{-5mm}
  \caption{ \label{fig:dns-points} \footnotesize
  The admissible domains of $\Delta$ for the Adler $D_\text{NS}$ function
under different series transformations.
Left: 1D case. The vertical dashed line denotes PMC,
$\Delta_0 = d_2[1]-\frac{11}{12}\approx -0.22$; 
the vertical solid line corresponds to BLM and gNNA,
$\Delta_0 = d_2[1]\approx 0.69\approx \ln(2)$, Eqs.(\ref{eq:d1-4}).
Right: 2D case. $\bullet$ is the asymptotic limit point $d_2=d_3=d_4=0$ at
$n_f=5:$ $\Delta_0 = 0.7352$, $\Delta_1 = -1.2097$ and $\Delta_2 = 2.8349$.
Different versions of PMC:
\black{ \ding{115} } \cite{Ma:2015dxa}, $(n_f=5(4))$, $\Delta_0\approx -0.22$
and $\Delta_1\approx 0.61\,(0.89)$, Eq.(\ref{eq:Delta});
\red{\ding{108}} \cite{Baikov:2022zvq}, $(n_f = 5)$, $\Delta_0 = 0.69$ and
$\Delta_1 = 8.59$;
\amethyst{\ding{117}} \cite{Goriachuk:2021ayq,Kataev:2022iqf}, $(n_f = 5)$,
$\Delta_0 = 0.69$ and $\Delta_1 = 8.586$.
gNNA: \BluTn{\ding{108}} \cite{Mikhailov:2024mrs}, $n_f = 5(4)$,
$\Delta_0\approx 0.69$ and $\Delta_1\approx 1.320\,(1.171)$.
}
\end{figure}

In a similar manner, the results for the Adler $D_\text{NS}$ have been obtained and are shown for the $D=1, 2$ cases with the same notation in Fig. \ref{fig:dns-points}.
Among various PMC predictions only the result~ \black{ \ding{115} } \cite{Ma:2015dxa} is acceptable again,
which we consider here  an accident.
The reason for this is that the condition $d_n[0] \sim \gamma_n^\text{photon}$,
which is used in this approach to construct the perturbation series for $D_\text{NS}$,
has no justification.

\subsection{What does gNNA mean?}
Let us consider the contributions that are generated by radiative corrections to a single gluon line in order $n$.
The sum of these corrections has been introduced as a generalization of naive non-Abelianization (gNNA),
replacing the single term $\beta_0^{n-1}c_n[n-1]$ of NNA \cite{Broadhurst:1994se} with this sum of terms,
as described here in Appendix \ref{App:A1}, in detail in \cite{Mikhailov:2024mrs},
and long ago in \cite{Bakulev:2010gm}.
Diagrammatically, this can be interpreted as the momentum flowing through this dressed gluon line in order n,
which shapes a new normalization scale $\mu'$ for the charge $a_s$.
In fixed order, there is a single momentum flow through the gluon line that is not suppressed by $1/\beta_0$ (if we consider $\beta_0$ as large parameter as discussed in \cite{Mikhailov:2024mrs}). This is similar to the interpretation of the BLM term $c_2[1]$ for $n = 2$.
The complement to the gNNA contribution at each order provides an alternative to the artificial ``conformal'' element $c_n[0]$,
replenishing them with the other (suppressed by $1/\beta_0$) terms. Therefore, one should take the complement of this ``flow'' contribution in each coefficient $c'_n$ of the PT
\ba
C' &=& 1 + c_1^\text{ns} \Big\{a'_s\,1 + (a'_s)^2 c_2[0] + (a'_s)^3 \left(\beta_0 c_3[1]+ c_3[0] \right)+ (a'_s)^4 \left(\beta_0^2 c_4[2]+\beta_1 c_4[0,1]+ \right. \nonumber \\
&&\left. \phantom{ 1 + c_1^\text{ns} \Big\{a'_s\,1 + (a'_s)^2 c_2[0] + (a'_s)^3 (\beta_0 c_3[1]+ c_3[0]} \beta_0 c_4[1]+c_4[0]\right)+ \ldots \Big\},
\ea
which corrects the PMC contributions $c_n[0]$ to each of the coefficients starting with $n=3$.
The parts of the rest in each order are transferred to the shifts $\Delta_i$ that  follow from Eq.(\ref{eq:rearrange}),
\begin{subequations}
  \label{eq:gNNA}
\ba
\beta_0\Delta_{0}&=&c_2-c_2[0]=\beta_0 c_2[1]; \\
\beta_0^2\Delta_{1}&=&c_3 - 2c_2\,\beta_0 c_2[1] - \frac{\beta_1}{\beta_0}\beta_0 c_2[1] + (\beta_0 c_2[1])^2 -(\beta_0c_3[1]+c_3[0]) \\
\beta_0^3\Delta_{2}&=&c_4 - 3 c_3\beta_0c_2[1] - c_2\left( 2 \beta_1c_2[1]-3 (\beta_0 c_2[1])^2+2 \beta_0^2\Delta_{1}\right)- (\beta_0 c_2[1])^3
 - \beta_2c_2[1]+\nonumber \\
&&\!\frac{5}{2}\beta_1\beta_0(c_2[1])^2\! -\!\Delta_1 \beta_0(\beta_1\!-\!\frac{5}{2}\beta_0^2c_2[1]) \!-\!\left(\beta_0^2c_4[2]\!+\!\beta_1c_4[0,1]\!+\!\beta_0c_4[1]\!+\!c_4[0]\right).
\ea
\end{subequations}
The sign \BluTn{\ding{108}}$(\!\Delta_0\! =\!2, \Delta_1\!\approx\!3.058)$ 
in Fig.\ref{fig:bsr-points} (see \BluTn{\ding{108}}$(\Delta_0\!\approx\!0.69, \Delta_1\!\approx\!1.32)$ in Fig.\ref{fig:dns-points})
means gNNA obtained from Eqs.(\ref{eq:gNNA}). It works at $t\geqslant 8 (\simeq 300$\,GeV$^2$).

\section{Conclusions }
\label{sec:sec6}
We have studied renormalization group invariant
(physical) quantities and constructed a modification of the
renormalization group transformation for the perturbation series by
introducing the parameterization for a new scale
$\mu'$, where $\mu \to \mu'$ and $a_s(\mu^2) \to a_s(\mu'^2)$.
To determine this scale $\mu'$ in the $n$-th order of the perturbation
series, we have introduced the parameters
$\Delta_0,\Delta_1,\ldots \Delta_{n-2}\equiv \{\Delta_i \}_0^{n-2}$.
The values of the parameters $\Delta_i$ are not arbitrary.
We proposed the necessary conditions on the admissible domain of
$\{\Delta_i \}$ in Section~\ref{sec:sec3}, which are based on the evident
hierarchy of perturbation series and, which we called filters.

As an example, we have implemented the construction of the filters
for $\{\Delta_i \}$ relative to the Bjorken polarized sum rule ($C_\text{Bjp}$) in DIS.
We discussed in Section~\ref{sec:sec4} the possible optimizations of the
truncated perturbation series for $C_\text{Bjp}$ with the help of the parameters
$\{\Delta_i \}$.
We have thus obtained the conditions for the truncations
$c_4=0$ and $c_3=c_4=0$ for the perturbation $C_\text{Bjp}$ coefficients.
Similar results for the Adler $D_\text{NS}$-function were
presented in Appendix\,\ref{App:B}.
Such modified perturbation series  for the $D_\text{NS}$ and $C_\text{Bjp}$ could be directly used in the
future to process the corresponding experimental data.

Furthermore, in Section~\ref{sec:sec5}, we have reviewed the results of
various techniques presented in the literature to transform perturbation
series of physical quantities, such as the $C_\text{Bjp}$ and Adler $D_\text{NS}$-function,
against the background of the previously introduced filters.
We found that only a few of these results met the filter criteria.

The important, but auxiliary, material on the $\{\beta\}$-expansion has
been placed in Appendices~\ref{App:A} and \ref{App:B}.

\begin{acknowledgments}
We would like to thank A.~G. Grozin and A.~L. Kataev
for  fruitful discussions.  
D. K. thanks A. Kotlorz for his help in numerical computations.
\end{acknowledgments}

\begin{appendix}
\appendix
\section{$\{\beta \}$-expansion for RGI quantity $C$}
 \renewcommand{\theequation}{\thesection.\arabic{equation}}
\label{App:A}   \setcounter{equation}{0}
\subsection{General form of the $\{\beta \}$-expansion}
 \label{App:A1}
The $\{\beta\}$-expansion representation introduced in \cite{Mikhailov:2004iq}
prescribes the following form of decomposition of the perturbation coefficients $c_n$ for $C_\text{Bjp}$ in
Eq.~(\ref{eq:4.1b}), or for any other RGI quantities $C = D_\text{NS}, R_{(e^-e^+\to h)},\ldots$.
This requests (i) obtaining a two-dimensional matrix $C_{n\bm{i}}$ instead of a series $c_n$ and
(ii) that each element of the matrix represent a unique contribution to the renormalization of the charge $a_s$ that is specific to that particular physical quantity. For $C=C_\text{Bjp}$ one has:
\begin{subequations}
\label{eq:d_beta}
\begin{eqnarray}
\label{eq:c_1}
\!\!\!\!\!\!\!\!\!\!c^\text{ns}_1\!\!&=&\!-3C_\text{F}\, , \\
\!\!\!\!\!\!\!\!\!\!c_2&=&\! \beta_0\,c_2[1]
  + c_2[0]\, ,\label{eq:c_2}\\
\!\!\!\!\!\!\!\!\!\!c_3
&=&\!
  \beta_0^2\,c_3[2]
  + \beta_1\,c_3[0,1]
  +       \beta_0 \,  c_3[1]
  + c_3[0]\, ,\label{eq:c_3} \\
\!\!\!\!\!\!\!\! c_4
   &=&\! \beta_0^3\, c_4[3]
     + \beta_1\,\beta_0\,c_4[1,1]
     + \beta_2\, c_4[0,0,1]
     +\! \beta_0^2\,c_4[2]
     +\! \beta_1  c_4[0,1]
     +\! \beta_0c_4[1]+c_4[0], \label{eq:c_4} \\
   &\vdots& \nonumber \\
\!\!\!\!\!\!\!\!\!\!c_{n}
   &=&\!\!\beta_0^{n-1}\, c_{n}[n\!-\!1]+ \cdots + \beta_0\,c_n[1]+ c_n[0]\,,
\label{eq:c_n}
\end{eqnarray}
\end{subequations}
where  $\beta_i$ are the expansion coefficients of the QCD $\beta$-function presented in Appendix \ref{App:C},
\begin{equation}
\label{eq:beta}
\mu^2\frac{d a_s(\mu^2)}{d \mu^2}=
\beta(a_s)=-a_s^{2}(\mu^2) \sum_{i\geq 1} \beta_{i-1} a_s^{i-1}(\mu^2)\,.
\end{equation}
The decomposition in Eqs.(\ref{eq:d_beta}) contains complete knowledge about
$\alpha_s$-renormalization in each order of expansion for the RGI quantity $C_\text{Bjp}$.
In the general case of order $n$, the rule for  appearance $\beta$-function coefficients
is considered in \cite{Mikhailov:2024mrs}.
NNA corresponds to considering $\beta_0^{n-1} c_{n}[n-1]$ in each order n, while gNNA means that all the elements $c_{n}[j_1, \ldots, j_{n-1}]$ with $j_1 + 2j_2 + \cdots + j_{n-1} = n-1$ are taken into account in this order.
The $\{\beta\}$-expansion allows us to work on any optimization goal for the perturbation series.

\subsection{ RGI quantity $C_\text{Bjp}$ and its $\{\beta \}$-expansion}
 \label{App:A2}
The explicit form of the elements of $\{\beta\}$-expansion in Eq.~(\ref{eq:d_beta}) (for $C_\text{Bjp}$)
can be based on the results obtained within the extended QCD (QCDe) in \cite{Chetyrkin:2022fqk}.
These elements have been confirmed later in diagrammatic analysis in \cite{Mikhailov:2024mrs}, which coincide with the result obtained
independently in  \cite{Kataev:2014jba,Baikov:2022zvq} (normalized on $c^\text{ns}_1$) reads
  \begin{subequations}
 \label{eq:c1-4}
\begin{eqnarray}
c_1&=&1;~~~~~c^\text{ns}_1=-3{\rm  C_F}; \label{c-11}\\
c_2[1]&=&  2;~~~~~
c_2[0]= \frac{1}{3}{\rm C_A}-\frac{7}{2}{\rm C_F}=-\frac{11}{3};
 \label{c-21} \\
c_3[2]&=& \frac{115}{18} ;~c_3[0,1]=\bigg(\frac{59}{12}-4\zeta_3\bigg)\approx 0.11;
\label{C-32} \\
c_3[1]&=& -\bigg[{\rm C_F}\bigg(\frac{166}{9}- \frac{16}3\zeta_3\bigg) +
{\rm C_A}\bigg(\frac{215}{36}- 32 \zeta_3+\frac{40}{3}\zeta_5\bigg)\bigg]\approx 39.96;
\label{C-31} \\
c_3[0] &=&{\rm C_A^2}
\bigg(\frac{523}{36} - 72\zeta_3\bigg)+\frac{65}{3}{\rm C_F C_A}+ \frac{\rm C_F^2}{2}\approx 560.62.
\label{C-30}
\end{eqnarray}
\end{subequations}
\begin{subequations}
 \label{eq:c_4expr}
 \begin{eqnarray}
 c_4[3] &=& \frac{605}{27};
  c_4[1,1]= \left(\frac{715}{24}- \frac{677}9 \zeta_3 +\frac{220}3 \zeta_5\right); \\
 c_4[2] &=&  C_\text{F} \left(-\frac{8057}{72} + 32  \zeta_3\right) +
C_\text{A} \left(-\frac{28615}{144 \cdot 3} + \frac{4105}{18}  \zeta_3 + 8  \zeta_3^2 - \frac{370}3  \zeta_5\right);    \\
  c_4[0,0,1]&=&  \left(\frac{146}9 + \frac{148}3 \zeta_3 - 80 \zeta_5\right); \\
 c_4[1] &=&  C_\text{F}^2\left(\frac{1478}9+\frac{824}3 \zeta_3-\frac{1520}3\zeta_5\right)- \nonumber  \\
 &&\phantom{C_\text{F} \bigg[ }C_\text{A} C_\text{F} \left(-\frac{1177}{36} + \frac{5888}9 \zeta_3 - \frac{2000}{9}
 \zeta_5 + 560 \zeta_7\right)- \nonumber  \\
 &&\phantom{C_\text{F} \bigg[ } C_\text{A}^2 \left(-\frac{3829}{72\cdot 3} + 762 \zeta_3 - \frac{6250}{9} \zeta_5 + 140
 \zeta_7\right); \\
  c_4[0,1]&=&  C_\text{A} \left(-\frac{109}{12} - \frac{1006}{9} \zeta_3 + \frac{460}{3} \zeta_5\right) +
    C_\text{F} \left(-\frac{1399}{36} + \frac{100}3 \zeta_3 \right);
 \end{eqnarray}
 \begin{eqnarray}
 c_4[0] &=& \tilde{c}_4[0]+ \delta c_4\nonumber \\
        &=&   C_\text{F}^3 \left(\frac{4823}{24} + 32 \zeta_3\right)
 -  C_\text{A} C_\text{F}^2 \left(\frac{1067}{2} + 144 \zeta_3\right)+
  C_\text{A}^2 C_\text{F}\left(\frac{685}{4} + 648 \zeta_3\right)- \nonumber\\
 &&  C_\text{A}^3 \left(-\frac{68047}{48\cdot 3} + \frac{8113}{6} \zeta_3 - 2370 \zeta_5\right) +  \delta c_4\,;
 \label{eq:c_40} \\
 \delta c_4=-\delta d_4&\!=\!&\!-\frac{16}{3 C_\text{F}}\Big[n_f \frac{d_F^{a b c d}d_F^{a b c d}}{d_F} (13 + 16 \zeta_3 - 40 \zeta_5) +
 \frac{d_A^{a b c d}d_F^{a b c d}}{d_F} (-3 + 4 \zeta_3 + 20\zeta_5)\!\Big]. \label{eq:c_40delta}
\end{eqnarray}
 \end{subequations}
with the $SU_{c}(N)$-group fundamental fermion invariants
\begin{eqnarray}
\label{eq:inv}
&&T_r = \frac{1}{2}\, ; \; C_\text{F}= \frac{N^2-1}{2N}\, ; \; C_A =N\, ;\;
N_A = 2C_\text{F} C_\text{A} \equiv N^2-1\, ; \nonumber \\
&&d^{abc}d^{abc}= \frac{(N^2-4)N_A}{N}\, ; ~\dRA = \frac{N(N^2+6)}{48}N_A\, ; \nonumber \\
&&\dRR = \frac{N^4-6N^2+18}{96N^2}N_A\, ; ~ \dAA = \frac{N^2(N^2+36)}{24}N_A\, ,
\end{eqnarray}
where $d_R$ is the dimension of the quark color representation, $d_F = 3$ in QCD, and $n_f$ denotes the number of active flavors.
The numerical form of $C_\text{Bjp}(a_s)$ \cite{Baikov:2010je} reads
\begin{eqnarray}
 \label{eq:C1-4numer}
C_\text{Bjp}(a_s)&=&1-4\Big[ a_s + a_s^2\left(\frac{55}3-\frac{4}3n_f\right)+a_s^3\left(663.04-121.72n_f+2.84n_f^2\right)+ \nonumber \\
&& a_s^4\left(30684.6 - 7897.05n_f + 482.64n_f^2 - 6.64n_f^3\right)\Big]\,.
\end{eqnarray}

An alternative  kind of $c_n[.]$ elements, presented in \cite{Goriachuk:2021ayq},
was obtained from the Adler function $D_\text{NS}$ elements
$d_k[.]$ by means of the CBK relation \cite{Broadhurst:1993ru} (see Appendix\,\ref{App:B}).
The other set of the  $c_n[.]$ elements appears in a similar way
using the CBK relation but is based on a different construction of the elements $d_k[.]$ -- ``QCD degeneracy relations'' -- in the PMC approach in \cite{Ma:2015dxa}.

\section{ RGI quantity $D_\text{NS}$ and its $\{\beta \}$-expansion}
 \label{App:B}
  \renewcommand{\theequation}{\thesection.\arabic{equation}}
   \setcounter{equation}{0}
   \vspace{-1mm}

\begin{subequations}
 \label{eq:c-MS}
For the default  condition $Q^2=\mu^2$,
 \ba
\!\!\!\!\! D_\text{NS}\left(a_s(\mu^2)\right)&=& d_F \left[
1+d^\text{ns}_1\Big(a_s(\mu^2)+d_2~a_s^2(\mu^2)+ d_3~a_s^3(\mu^2)+d_4~a_s^4(\mu^2)+\ldots \Big) \right], \label{eq:D0} \\
d^\text{ns}_1&=&3{\rm C_F}=4~ (\text{at}\, N_c=3), \label{D-11}
\ea
and in numerical form
\ba
D_\text{NS}(a_s)&=&d_F\bigg\{1 + 4 a_s \Big[1 + a_s \left(7.9428 - 0.4612 n_f \right) +
    a_s^2 \left(291.8832 - 67.4528 n_f + 1.3792 n_f^2\right) +\nonumber \\
    && a_s^3 \left(8690.6624 - 2204.1728 n_f + 120.0192 n_f^2 - 0.6464 n_f^3\right)\Big]
    \bigg\}.
\ea
 \end{subequations}

Substituting the conditions $d_3 = d_4 = 0$ into Eq.(\ref{eq:c4=c3=0}), which is written for $D_{\text{NS}}$,
results in a moderate shift of $\bm{\Delta}$. Specifically,
we obtain $\Delta^{(5, (4))}_0 = -0.359 (-0.375)$ and $\Delta^{(5, (4))}_1 = 0.708 (1.604)$:
\begin{subequations}
 \label{eq:d3=d4=0}
 \ba
 D\left(t,\{\Delta \}\right)&=&
1+d_1^\text{ns}\Big(1\,a'_s+d'_2(\Delta_0 )\,a_s'^2 + 0 + 0\Big), \label{eq:B3.a}\\
 \text{where,}&& d_2'(\Delta_0 )= d_2-\beta_0\cdot \Delta^{(4,5)}_0 \stackrel{n_f=5}{\longrightarrow}\frac{1}{3}+\beta_0\cdot 0.69 +\beta_0\cdot 0.359
 ;        \label{eq:B3.b}   \\
&& a_s' = \bar{a}_s(t')=\bar{a}_s(t - \Delta(a'_s));~ \Delta(a'_s)=\Delta^{(4,5)}_0+a_s'\Delta^{(4,5)}_1\,. \label{eq:B3.c}
  \ea
   \end{subequations}

For the Adler function $D_\text{NS}$ the decomposition of the elements up to order $O(a^3)$ corresponding to Eq.(\ref{eq:d_beta}) reads
\cite{Mikhailov:2004iq,Kataev:2014jba}
\begin{subequations}
\label{eq:d1-4}
 \begin{eqnarray}
d_2[1]&=&\frac{11}2-4\zeta_3;~
d_2[0]=\frac{\rm C_A}3-\frac{\rm C_F}2; \label{D-21} \\
d_3[2]&=&\frac{302}9-\frac{76}3\zeta_3;~d_3[0,1]=\frac{101}{12}-8\zeta_3;\label{D-32}\\
d_3[1]&=&
    {\rm C_A}\left(-\frac{3}4 + \frac{80}3\zeta_3 -\frac{40}3\zeta_5\right) -
    {\rm C_F}\left(18 + 52\zeta_3 - 80\zeta_5\right) \label{D-31}]; \label{eq:d31} \\
d_3[0]&=& {\rm C_A^2}\left(\frac{523}{36}- 72 \zeta_3\right)
    +\frac{71}3 {\rm C_A C_F} - \frac{23}{2} {\rm C_F^2}~.  \label{D-30}
\end{eqnarray}
\end{subequations}
The results in Eqs.(\ref{eq:d1-4}) were confirmed in \cite{Baikov:2022zvq} based on the results obtained within QCDe
\cite{Chetyrkin:2022fqk} and in \cite{Mikhailov:2024mrs}.
We do not present here the elements $d_4[.]$ explicitly, referring reader to the corresponding references in \cite{Baikov:2022zvq,Mikhailov:2024mrs}.
The results of alternative approaches to the $\{\beta \}$-expansion for $D_\text{NS}, C_\text{Bjp}$
are discussed in detail in \cite{Kataev:2023xqr}.

\section{ RG solutions for QCD charge}
 \label{App:C}
\textbf{1.}
Asymptotic freedom is  the basic
feature of QCD as a
theory of strong interactions \cite{Gross:1973id,Politzer:1973fx}.
This leading
order prediction was quickly complemented by the corresponding
two-loop \cite{Caswell:1974gg,Jones:1974mm},  three-loop
\cite{Tarasov:1980au,Larin:1993tp} results.
The  four-loop result was obtained seventeen  years later \cite{vanRitbergen:1997va} and
here we stay at this level of accuracy.
The explicit expressions for the first coefficients of $\beta$ function expansion are
\begin{subequations}
 \begin{eqnarray}
    \beta_0 &=& \frac{11}{3}\,C_\text{A} - \frac{4}{3}\,T_r n_f
    \,;\qquad
    \beta_1 = \frac{34}{3}\,C_{\text{A}}^{2}
        - \left(4C_\text{F}
        + \frac{20}{3}\,C_\text{A}\right)T_r n_f ; \\
    \beta_2 &=&   \frac{2857}{54} C_A^3
 +2 C_F^2 T_r n_f - \frac{205}{9} C_F C_A T_r n_f
 - \frac{1415}{27} C_A^2 T_r n_f
 + \frac{44}{9} C_F (T_r n_f)^2 \nonumber \\
 && + \frac{158}{27} C_A (T_r n_f)^2\, ;
\end{eqnarray}
\begin{eqnarray}
\beta_3 &=&
C_A^4 \left( \frac{150653}{486} - \frac{44}{9} \zeta_3 \right)+
C_A^3 T_R n_f \left( - \frac{39143}{81} + \frac{136}{3} \zeta_3 \right)
+ C_F^2 T_R^2 n_f^2 \left( \frac{1352}{27} - \frac{704}{9} \zeta_3
\right) \nonumber \\
 && +C_A C_F T_R^2 n_f^2 \left( \frac{17152}{243} + \frac{448}{9} \zeta_3 \right)
 + C_A C_F^2 T_R n_f \left( - \frac{4204}{27} + \frac{352}{9} \zeta_3 \right) + \frac{424}{243} C_A
 T_R^3 n_f^3 \nonumber \\
&&+ C_A^2 C_F T_R n_f \left( \frac{7073}{243} - \frac{656}{9} \zeta_3 \right)
+ C_A^2 T_R^2 n_f^2 \left( \frac{7930}{81} + \frac{224}{9} \zeta_3
\right) + \frac{1232}{243} C_F T_R^3 n_f^3 \nonumber \\
&&+ 46 C_F^3 T_R n_f + n_f \dRANA \left( \frac{512}{9} - \frac{1664}{3} \zeta_3 \right) + n_f^2
\dRRNA \left( - \frac{704}{9} + \frac{512}{3} \zeta_3
\right) \nonumber \\
&&+ \dAANA \left( - \frac{80}{9} + \frac{704}{3} \zeta_3 \right).
 \label{eq:beta0&1&2}
\end{eqnarray}
 \end{subequations}
The corresponding four-loop RG equations for the coupling
$A=\beta_0\,\alpha_s/(4\pi)$ read
\begin{eqnarray}
\!\!\!\! \frac{d A_{(4)}}{dt}
  = - A_{(4)}^2\left[1 + b_1\,A_{(4)}+ b_2\,A_{(4)}^2+b_3\,A_{(4)}^3\right]
 \text{~~with~}b_i
 \equiv
 \frac{\beta_i}{\beta_0^{i+1}}\,, t=\ln\left(Q^2/\Lambda^2_{qcd} \right).
\label{eq:beta3.new}
\end{eqnarray}
 The approximate solution of the renormalization-group equation in the four-loop
QCD \cite{Tanabashi:2018}, where the $\beta$-function is given by
Eq.~(\ref{eq:beta3.new}), assumes the asymptotic expansion
 \begin{eqnarray}
 A_{(4)}(t) &\simeq& \frac{1}{t} \, \Big[ \,
1-\frac{b_1 l}{t} + \frac{1}{t^2}\Big( b_1^2 (l^2-l-1) + b_2 \Big) \Big. \nonumber \\
&&
+ \frac{1}{2t^3}\Big( b_1^3 (-2l^3+5l^2+4l-1) - 6b_1b_2l + b_3 \Big)
+ \frac{1}{6t^4} \Big( b_1^2b_2 (2l^2-l-1) \Big. \nonumber \\
&&
\Big.\Big. + b_1^4 (6l^4-26l^3-9l^2+24l+7) - b_1b_3 (12l+1) + 10 b_2^2 \Big) \, \Big] \, ,
\label{eq:beta.new.4L}
\end{eqnarray}
where $t=\ln(Q^2/\Lambda_{(4)}^2);~  l=\ln (t)$.

\begin{figure}[bh]
\centerline{\includegraphics[width=0.8\textwidth]{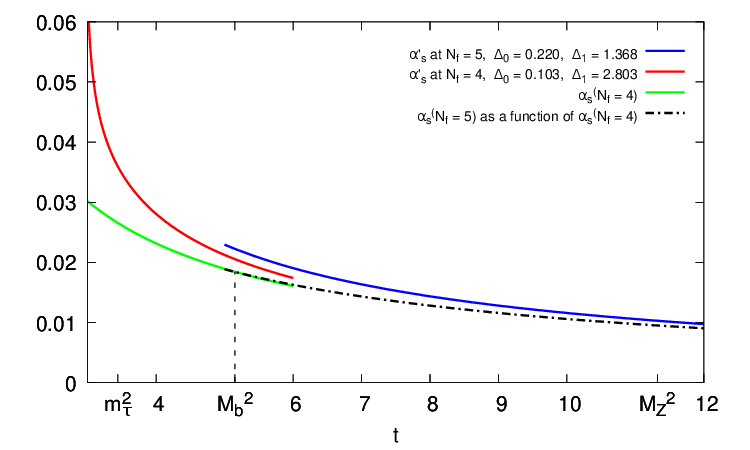}} \vspace{-4mm}
\caption{\!\!\!\! \label{fig:alpha-mod} \footnotesize
Comparison of $a'_s(t)$, Eq.(\ref{eq:a'_s}), and $a_s(n_f=4,5)$.
The solid green line shows $a_s(n_f=4,t)$; the dash-dotted black line denotes $a_s(n_f=5,t)$ that corresponds to equation (\ref{eq:C4a}).
The solid blue (red) lines are  $a'(n_f=5,t) (a'(n_f=4,t))$, whose argument  $t$ is shifted  in $\{ \bm{\Delta}\}$  in
accordance with the condition $c_3=c_4=0$ in equation (\ref{eq:4.3d}).
 }
\end{figure}

Based on the matching condition \cite{Chetyrkin:2005ia} on the threshold $m=M_b$ 
-- the scale-invariant mass of the bottom quark -- for $a_{(f)} \to a_{(f+1)}$, 
one obtains
\begin{subequations}
 \label{eq:C4}
 \begin{eqnarray}
a_{(f+1)}(t)=a_{(f)}(t) &&\left\{
1+ \frac{2}{3}a_{(f)}(t) L + a_{(f)}^2\left[\left(\frac{2}{3}L\right)^2 +\frac{38}{3}L - \frac{22}{9}\right]+ a_{(f)}^3(t) \right.
\nonumber \\
  &&\left. \left[\left(\frac{2}{3}L\right)^3 +\frac{511}{9}L^2+\frac{2191}{9}L +C_1+C_2 \right]   \right\}, \label{eq:C4a}
\end{eqnarray}
where
\begin{eqnarray}
L=t-\ln\left(m^2/\Lambda_{(4)}^2\right)= \ln\left(Q^2/m^2\right),  \\
C_1=\left( \frac{82043}{432} \zeta_3- \frac{564731}{1944}\right);~C_2=4 \left( \frac{281}{27}L- \frac{2633}{486}\right)
\end{eqnarray}
 \end{subequations}
 The expressions in Eqs.(\ref{eq:beta.new.4L},\ref{eq:C4}) we use in our numerical calculations in Secs.\ref{sec:sec4},\,\ref{sec:sec5}, Appendix\,\ref{App:B}.
In Fig.~\ref{fig:alpha-mod} we compare the behavior of $a'_s(t)$ in Eq.(\ref{eq:a'_s})
and $a_s(n_f=4,5)$.
\end{appendix}
\bibliographystyle{JHEP}

\begin{thebibliography}{34}
\expandafter\ifx\csname natexlab\endcsname\relax\def\natexlab#1{#1}\fi
\expandafter\ifx\csname bibnamefont\endcsname\relax
  \def\bibnamefont#1{#1}\fi
\expandafter\ifx\csname bibfnamefont\endcsname\relax
  \def\bibfnamefont#1{#1}\fi
\expandafter\ifx\csname citenamefont\endcsname\relax
  \def\citenamefont#1{#1}\fi
\expandafter\ifx\csname url\endcsname\relax
  \def\url#1{\texttt{#1}}\fi
\expandafter\ifx\csname urlprefix\endcsname\relax\def\urlprefix{URL }\fi
\providecommand{\bibinfo}[2]{#2}
\providecommand{\eprint}[2][]{\url{#2}}

\bibitem[{\citenamefont{Kotlorz and Mikhailov}(2019)}]{Kotlorz:2018bxp}
\bibinfo{author}{\bibfnamefont{D.}~\bibnamefont{Kotlorz}} \bibnamefont{and}
  \bibinfo{author}{\bibfnamefont{S.~V.} \bibnamefont{Mikhailov}},
  \bibinfo{journal}{Phys. Rev. D} \textbf{\bibinfo{volume}{100}},
  \bibinfo{pages}{056007} (\bibinfo{year}{2019}), \eprint{1810.02973}.

\bibitem[{\citenamefont{Baikov et~al.}(2008)\citenamefont{Baikov, Chetyrkin,
  and Kuhn}}]{Baikov:2008jh}
\bibinfo{author}{\bibfnamefont{P.~A.} \bibnamefont{Baikov}},
  \bibinfo{author}{\bibfnamefont{K.~G.} \bibnamefont{Chetyrkin}},
  \bibnamefont{and} \bibinfo{author}{\bibfnamefont{J.~H.} \bibnamefont{Kuhn}},
  \bibinfo{journal}{Phys. Rev. Lett.} \textbf{\bibinfo{volume}{101}},
  \bibinfo{pages}{012002} (\bibinfo{year}{2008}), \eprint{0801.1821}.

\bibitem[{\citenamefont{Baikov et~al.}(2010)\citenamefont{Baikov, Chetyrkin,
  and Kuhn}}]{Baikov:2010je}
\bibinfo{author}{\bibfnamefont{P.~A.} \bibnamefont{Baikov}},
  \bibinfo{author}{\bibfnamefont{K.~G.} \bibnamefont{Chetyrkin}},
  \bibnamefont{and} \bibinfo{author}{\bibfnamefont{J.~H.} \bibnamefont{Kuhn}},
  \bibinfo{journal}{Phys. Rev. Lett.} \textbf{\bibinfo{volume}{104}},
  \bibinfo{pages}{132004} (\bibinfo{year}{2010}), \eprint{1001.3606}.

\bibitem[{\citenamefont{Baikov et~al.}(2012)\citenamefont{Baikov, Chetyrkin,
  Kuhn, and Rittinger}}]{Baikov:2012zm}
\bibinfo{author}{\bibfnamefont{P.~A.} \bibnamefont{Baikov}},
  \bibinfo{author}{\bibfnamefont{K.~G.} \bibnamefont{Chetyrkin}},
  \bibinfo{author}{\bibfnamefont{J.~H.} \bibnamefont{Kuhn}}, \bibnamefont{and}
  \bibinfo{author}{\bibfnamefont{J.}~\bibnamefont{Rittinger}},
  \bibinfo{journal}{JHEP} \textbf{\bibinfo{volume}{07}}, \bibinfo{pages}{017}
  (\bibinfo{year}{2012}), \eprint{1206.1284}.

\bibitem[{\citenamefont{Brodsky et~al.}(1983)\citenamefont{Brodsky, Lepage, and
  Mackenzie}}]{Brodsky:1982gc}
\bibinfo{author}{\bibfnamefont{S.~J.} \bibnamefont{Brodsky}},
  \bibinfo{author}{\bibfnamefont{G.~P.} \bibnamefont{Lepage}},
  \bibnamefont{and} \bibinfo{author}{\bibfnamefont{P.~B.}
  \bibnamefont{Mackenzie}}, \bibinfo{journal}{Phys. Rev.}
  \textbf{\bibinfo{volume}{D28}}, \bibinfo{pages}{228} (\bibinfo{year}{1983}).

\bibitem[{\citenamefont{Mikhailov}(2007)}]{Mikhailov:2004iq}
\bibinfo{author}{\bibfnamefont{S.~V.} \bibnamefont{Mikhailov}},
  \bibinfo{journal}{JHEP} \textbf{\bibinfo{volume}{06}}, \bibinfo{pages}{009}
  (\bibinfo{year}{2007}), \eprint{hep-ph/0411397}.

\bibitem[{\citenamefont{Kataev and Mikhailov}(2015)}]{Kataev:2014jba}
\bibinfo{author}{\bibfnamefont{A.~L.} \bibnamefont{Kataev}} \bibnamefont{and}
  \bibinfo{author}{\bibfnamefont{S.~V.} \bibnamefont{Mikhailov}},
  \bibinfo{journal}{Phys. Rev.} \textbf{\bibinfo{volume}{D91}},
  \bibinfo{pages}{014007} (\bibinfo{year}{2015}), \eprint{1408.0122}.

\bibitem[{\citenamefont{Baikov and Mikhailov}(2022)}]{Baikov:2022zvq}
\bibinfo{author}{\bibfnamefont{P.~A.} \bibnamefont{Baikov}} \bibnamefont{and}
  \bibinfo{author}{\bibfnamefont{S.~V.} \bibnamefont{Mikhailov}},
  \bibinfo{journal}{JHEP} \textbf{\bibinfo{volume}{09}}, \bibinfo{pages}{185}
  (\bibinfo{year}{2022}), \eprint{2206.14063}.

\bibitem[{\citenamefont{Mikhailov}(2024)}]{Mikhailov:2024mrs}
\bibinfo{author}{\bibfnamefont{S.~V.} \bibnamefont{Mikhailov}},
  \bibinfo{journal}{JHEP} \textbf{\bibinfo{volume}{10}}, \bibinfo{pages}{166}
  (\bibinfo{year}{2024}), \eprint{2406.15014}.

\bibitem[{\citenamefont{Mojaza et~al.}(2013)\citenamefont{Mojaza, Brodsky, and
  Wu}}]{Mojaza:2012mf}
\bibinfo{author}{\bibfnamefont{M.}~\bibnamefont{Mojaza}},
  \bibinfo{author}{\bibfnamefont{S.~J.} \bibnamefont{Brodsky}},
  \bibnamefont{and} \bibinfo{author}{\bibfnamefont{X.-G.} \bibnamefont{Wu}},
  \bibinfo{journal}{Phys. Rev. Lett.} \textbf{\bibinfo{volume}{110}},
  \bibinfo{pages}{192001} (\bibinfo{year}{2013}), \eprint{1212.0049}.

\bibitem[{\citenamefont{Di~Giustino et~al.}(2024)\citenamefont{Di~Giustino,
  Brodsky, Ratcliffe, Wu, and Wang}}]{DiGiustino:2023jiq}
\bibinfo{author}{\bibfnamefont{L.}~\bibnamefont{Di~Giustino}},
  \bibinfo{author}{\bibfnamefont{S.~J.} \bibnamefont{Brodsky}},
  \bibinfo{author}{\bibfnamefont{P.~G.} \bibnamefont{Ratcliffe}},
  \bibinfo{author}{\bibfnamefont{X.-G.} \bibnamefont{Wu}}, \bibnamefont{and}
  \bibinfo{author}{\bibfnamefont{S.-Q.} \bibnamefont{Wang}},
  \bibinfo{journal}{Prog. Part. Nucl. Phys.} \textbf{\bibinfo{volume}{135}},
  \bibinfo{pages}{104092} (\bibinfo{year}{2024}), \eprint{2307.03951}.

\bibitem[{\citenamefont{Cvetič and Kataev}(2016)}]{Cvetic:2016rot}
\bibinfo{author}{\bibfnamefont{G.}~\bibnamefont{Cvetič}} \bibnamefont{and}
  \bibinfo{author}{\bibfnamefont{A.~L.} \bibnamefont{Kataev}},
  \bibinfo{journal}{Phys. Rev.} \textbf{\bibinfo{volume}{D94}},
  \bibinfo{pages}{014006} (\bibinfo{year}{2016}), \eprint{1604.00509}.

\bibitem[{\citenamefont{Kataev and
  Molokoedov}(2023{\natexlab{a}})}]{Kataev:2022iqf}
\bibinfo{author}{\bibfnamefont{A.~L.} \bibnamefont{Kataev}} \bibnamefont{and}
  \bibinfo{author}{\bibfnamefont{V.~S.} \bibnamefont{Molokoedov}},
  \bibinfo{journal}{Phys. Part. Nucl.} \textbf{\bibinfo{volume}{54}},
  \bibinfo{pages}{931} (\bibinfo{year}{2023}{\natexlab{a}}),
  \eprint{2211.10242}.

\bibitem[{\citenamefont{Broadhurst and Kataev}(1993)}]{Broadhurst:1993ru}
\bibinfo{author}{\bibfnamefont{D.~J.} \bibnamefont{Broadhurst}}
  \bibnamefont{and} \bibinfo{author}{\bibfnamefont{A.~L.}
  \bibnamefont{Kataev}}, \bibinfo{journal}{Phys. Lett. B}
  \textbf{\bibinfo{volume}{315}}, \bibinfo{pages}{179} (\bibinfo{year}{1993}),
  \eprint{hep-ph/9308274}.

\bibitem[{\citenamefont{Grunberg and Kataev}(1992)}]{Grunberg:1991ac}
\bibinfo{author}{\bibfnamefont{G.}~\bibnamefont{Grunberg}} \bibnamefont{and}
  \bibinfo{author}{\bibfnamefont{A.~L.} \bibnamefont{Kataev}},
  \bibinfo{journal}{Phys. Lett. B} \textbf{\bibinfo{volume}{279}},
  \bibinfo{pages}{352} (\bibinfo{year}{1992}).

\bibitem[{\citenamefont{Deur et~al.}(2007)\citenamefont{Deur, Burkert, Chen,
  and Korsch}}]{Deur:2005cf}
\bibinfo{author}{\bibfnamefont{A.}~\bibnamefont{Deur}},
  \bibinfo{author}{\bibfnamefont{V.}~\bibnamefont{Burkert}},
  \bibinfo{author}{\bibfnamefont{J.-P.} \bibnamefont{Chen}}, \bibnamefont{and}
  \bibinfo{author}{\bibfnamefont{W.}~\bibnamefont{Korsch}},
  \bibinfo{journal}{Phys. Lett.} \textbf{\bibinfo{volume}{B650}},
  \bibinfo{pages}{244} (\bibinfo{year}{2007}), \eprint{hep-ph/0509113}.

\bibitem[{\citenamefont{Deur et~al.}(2016)\citenamefont{Deur, Brodsky, and
  de~Teramond}}]{Deur:2016cxb}
\bibinfo{author}{\bibfnamefont{A.}~\bibnamefont{Deur}},
  \bibinfo{author}{\bibfnamefont{S.~J.} \bibnamefont{Brodsky}},
  \bibnamefont{and} \bibinfo{author}{\bibfnamefont{G.~F.}
  \bibnamefont{de~Teramond}}, \bibinfo{journal}{Phys. Lett. B}
  \textbf{\bibinfo{volume}{757}}, \bibinfo{pages}{275} (\bibinfo{year}{2016}),
  \eprint{1601.06568}.

\bibitem[{\citenamefont{Ma et~al.}(2015)\citenamefont{Ma, Wu, Ma, Brodsky, and
  Mojaza}}]{Ma:2015dxa}
\bibinfo{author}{\bibfnamefont{H.-H.} \bibnamefont{Ma}},
  \bibinfo{author}{\bibfnamefont{X.-G.} \bibnamefont{Wu}},
  \bibinfo{author}{\bibfnamefont{Y.}~\bibnamefont{Ma}},
  \bibinfo{author}{\bibfnamefont{S.~J.} \bibnamefont{Brodsky}},
  \bibnamefont{and} \bibinfo{author}{\bibfnamefont{M.}~\bibnamefont{Mojaza}},
  \bibinfo{journal}{Phys. Rev.} \textbf{\bibinfo{volume}{D91}},
  \bibinfo{pages}{094028} (\bibinfo{year}{2015}), \eprint{1504.01260}.

\bibitem[{\citenamefont{Wu et~al.}(2013)\citenamefont{Wu, Brodsky, and
  Mojaza}}]{Wu:2013ei}
\bibinfo{author}{\bibfnamefont{X.-G.} \bibnamefont{Wu}},
  \bibinfo{author}{\bibfnamefont{S.~J.} \bibnamefont{Brodsky}},
  \bibnamefont{and} \bibinfo{author}{\bibfnamefont{M.}~\bibnamefont{Mojaza}},
  \bibinfo{journal}{Prog. Part. Nucl. Phys.} \textbf{\bibinfo{volume}{72}},
  \bibinfo{pages}{44} (\bibinfo{year}{2013}), \eprint{1302.0599}.

\bibitem[{\citenamefont{Goriachuk et~al.}(2022)\citenamefont{Goriachuk, Kataev,
  and Molokoedov}}]{Goriachuk:2021ayq}
\bibinfo{author}{\bibfnamefont{I.~O.} \bibnamefont{Goriachuk}},
  \bibinfo{author}{\bibfnamefont{A.~L.} \bibnamefont{Kataev}},
  \bibnamefont{and} \bibinfo{author}{\bibfnamefont{V.~S.}
  \bibnamefont{Molokoedov}}, \bibinfo{journal}{JHEP}
  \textbf{\bibinfo{volume}{05}}, \bibinfo{pages}{028} (\bibinfo{year}{2022}),
  \eprint{2111.12060}.

\bibitem[{\citenamefont{Deur et~al.}(2017)\citenamefont{Deur, Shen, Wu,
  Brodsky, and de~Teramond}}]{Deur:2017cvd}
\bibinfo{author}{\bibfnamefont{A.}~\bibnamefont{Deur}},
  \bibinfo{author}{\bibfnamefont{J.-M.} \bibnamefont{Shen}},
  \bibinfo{author}{\bibfnamefont{X.-G.} \bibnamefont{Wu}},
  \bibinfo{author}{\bibfnamefont{S.~J.} \bibnamefont{Brodsky}},
  \bibnamefont{and} \bibinfo{author}{\bibfnamefont{G.~F.}
  \bibnamefont{de~Teramond}}, \bibinfo{journal}{Phys. Lett.}
  \textbf{\bibinfo{volume}{B773}}, \bibinfo{pages}{98} (\bibinfo{year}{2017}),
  \eprint{1705.02384}.

\bibitem[{\citenamefont{Broadhurst and Grozin}(1995)}]{Broadhurst:1994se}
\bibinfo{author}{\bibfnamefont{D.~J.} \bibnamefont{Broadhurst}}
  \bibnamefont{and} \bibinfo{author}{\bibfnamefont{A.~G.}
  \bibnamefont{Grozin}}, \bibinfo{journal}{Phys. Rev. D}
  \textbf{\bibinfo{volume}{52}}, \bibinfo{pages}{4082} (\bibinfo{year}{1995}),
  \eprint{hep-ph/9410240}.

\bibitem[{\citenamefont{Bakulev et~al.}(2010)\citenamefont{Bakulev, Mikhailov,
  and Stefanis}}]{Bakulev:2010gm}
\bibinfo{author}{\bibfnamefont{A.~P.} \bibnamefont{Bakulev}},
  \bibinfo{author}{\bibfnamefont{S.~V.} \bibnamefont{Mikhailov}},
  \bibnamefont{and} \bibinfo{author}{\bibfnamefont{N.~G.}
  \bibnamefont{Stefanis}}, \bibinfo{journal}{JHEP}
  \textbf{\bibinfo{volume}{06}}, \bibinfo{pages}{085} (\bibinfo{year}{2010}),
  \eprint{1004.4125}.

\bibitem[{\citenamefont{Chetyrkin}(2022)}]{Chetyrkin:2022fqk}
\bibinfo{author}{\bibfnamefont{K.~G.} \bibnamefont{Chetyrkin}},
  \bibinfo{journal}{Nucl. Phys. B} \textbf{\bibinfo{volume}{985}},
  \bibinfo{pages}{115988} (\bibinfo{year}{2022}), \eprint{2206.12948}.

\bibitem[{\citenamefont{Kataev and
  Molokoedov}(2023{\natexlab{b}})}]{Kataev:2023xqr}
\bibinfo{author}{\bibfnamefont{A.~L.} \bibnamefont{Kataev}} \bibnamefont{and}
  \bibinfo{author}{\bibfnamefont{V.~S.} \bibnamefont{Molokoedov}},
  \bibinfo{journal}{Phys. Rev. D} \textbf{\bibinfo{volume}{108}},
  \bibinfo{pages}{096027} (\bibinfo{year}{2023}{\natexlab{b}}),
  \eprint{2309.03994}.

\bibitem[{\citenamefont{Gross and Wilczek}(1973)}]{Gross:1973id}
\bibinfo{author}{\bibfnamefont{D.~J.} \bibnamefont{Gross}} \bibnamefont{and}
  \bibinfo{author}{\bibfnamefont{F.}~\bibnamefont{Wilczek}},
  \bibinfo{journal}{Phys. Rev. Lett.} \textbf{\bibinfo{volume}{30}},
  \bibinfo{pages}{1343} (\bibinfo{year}{1973}).

\bibitem[{\citenamefont{Politzer}(1973)}]{Politzer:1973fx}
\bibinfo{author}{\bibfnamefont{H.~D.} \bibnamefont{Politzer}},
  \bibinfo{journal}{Phys. Rev. Lett.} \textbf{\bibinfo{volume}{30}},
  \bibinfo{pages}{1346} (\bibinfo{year}{1973}).

\bibitem[{\citenamefont{Caswell}(1974)}]{Caswell:1974gg}
\bibinfo{author}{\bibfnamefont{W.~E.} \bibnamefont{Caswell}},
  \bibinfo{journal}{Phys. Rev. Lett.} \textbf{\bibinfo{volume}{33}},
  \bibinfo{pages}{244} (\bibinfo{year}{1974}).

\bibitem[{\citenamefont{Jones}(1974)}]{Jones:1974mm}
\bibinfo{author}{\bibfnamefont{D.~R.~T.} \bibnamefont{Jones}},
  \bibinfo{journal}{Nucl. Phys.} \textbf{\bibinfo{volume}{B75}},
  \bibinfo{pages}{531} (\bibinfo{year}{1974}).

\bibitem[{\citenamefont{Tarasov et~al.}(1980)\citenamefont{Tarasov, Vladimirov,
  and Zharkov}}]{Tarasov:1980au}
\bibinfo{author}{\bibfnamefont{O.~V.} \bibnamefont{Tarasov}},
  \bibinfo{author}{\bibfnamefont{A.~A.} \bibnamefont{Vladimirov}},
  \bibnamefont{and} \bibinfo{author}{\bibfnamefont{A.~{\relax Yu}.}
  \bibnamefont{Zharkov}}, \bibinfo{journal}{Phys. Lett.}
  \textbf{\bibinfo{volume}{93B}}, \bibinfo{pages}{429} (\bibinfo{year}{1980}).

\bibitem[{\citenamefont{Larin and Vermaseren}(1993)}]{Larin:1993tp}
\bibinfo{author}{\bibfnamefont{S.~A.} \bibnamefont{Larin}} \bibnamefont{and}
  \bibinfo{author}{\bibfnamefont{J.~A.~M.} \bibnamefont{Vermaseren}},
  \bibinfo{journal}{Phys. Lett.} \textbf{\bibinfo{volume}{B303}},
  \bibinfo{pages}{334} (\bibinfo{year}{1993}), \eprint{hep-ph/9302208}.

\bibitem[{\citenamefont{van Ritbergen et~al.}(1997)\citenamefont{van Ritbergen,
  Vermaseren, and Larin}}]{vanRitbergen:1997va}
\bibinfo{author}{\bibfnamefont{T.}~\bibnamefont{van Ritbergen}},
  \bibinfo{author}{\bibfnamefont{J.~A.~M.} \bibnamefont{Vermaseren}},
  \bibnamefont{and} \bibinfo{author}{\bibfnamefont{S.~A.} \bibnamefont{Larin}},
  \bibinfo{journal}{Phys. Lett.} \textbf{\bibinfo{volume}{B400}},
  \bibinfo{pages}{379} (\bibinfo{year}{1997}), \eprint{hep-ph/9701390}.

\bibitem[{\citenamefont{Tanabashi et~al.}(2018)}]{Tanabashi:2018}
\bibinfo{author}{\bibfnamefont{M.}~\bibnamefont{Tanabashi}}
  \bibnamefont{et~al.} (\bibinfo{collaboration}{Particle Data Group}),
  \bibinfo{journal}{Phys. Rev.} \textbf{\bibinfo{volume}{D98}},
  \bibinfo{pages}{030001} (\bibinfo{year}{2018}).

\bibitem[{\citenamefont{Chetyrkin et~al.}(2006)\citenamefont{Chetyrkin, Kuhn,
  and Sturm}}]{Chetyrkin:2005ia}
\bibinfo{author}{\bibfnamefont{K.~G.} \bibnamefont{Chetyrkin}},
  \bibinfo{author}{\bibfnamefont{J.~H.} \bibnamefont{Kuhn}}, \bibnamefont{and}
  \bibinfo{author}{\bibfnamefont{C.}~\bibnamefont{Sturm}},
  \bibinfo{journal}{Nucl. Phys. B} \textbf{\bibinfo{volume}{744}},
  \bibinfo{pages}{121} (\bibinfo{year}{2006}), \eprint{hep-ph/0512060}.
\end{thebibliography}

\end{document}